\documentclass[article,10pt,nofootinbib,notitlepage,twocolumn]{revtex4-1}
\usepackage{graphicx}
\usepackage{amsmath}
\usepackage{amssymb}
\usepackage{dcolumn}
\usepackage{bm}
\usepackage{color}
\usepackage{dsfont}
\usepackage{feynmp}
\usepackage{marvosym}
\usepackage{phaistos}

\newcommand \beq{\begin{eqnarray}}
\newcommand \eeq{\end{eqnarray}}

\begin{document}
\unitlength=1mm
\allowdisplaybreaks

\title{Gauge fixing and physical symmetries}

\author{Duifje Maria van Egmond}
\affiliation{Centre de Physique Th\'eorique, CNRS, Ecole polytechnique, IP Paris, F-91128 Palaiseau, France.}

\author{Urko Reinosa}
\affiliation{Centre de Physique Th\'eorique, CNRS, Ecole polytechnique, IP Paris, F-91128 Palaiseau, France.}

\date{\today}

\begin{abstract}
We analyze how gauge fixing, required by any practical continuum approach to gauge systems, can interfere with the physical symmetries of such systems. In principle, the gauge fixing procedure, which deals with the (unphysical) gauge symmetry, should not interfere with the other (physical) symmetries. In practice, however, there can be an interference which takes two different forms.  First, depending on the considered gauge and on the considered physical symmetry, it might not always be simple or possible to devise approximation schemes that preserve the symmetry constraints on (gauge-independent) observables. Second, even at an exact level of discussion, the (gauge-dependent) effective action for the gauge field, and thus the related vertex functions, may not reflect the physical symmetries of the problem. We illustrate these difficulties using a very general class of gauge fixings that contains the usual gauge fixings as particular cases. Using background field techniques, we then propose specific gauge choices that allow one to keep the physical symmetries explicit, both at the level of the observables and at the level of the effective action for the gauge field. Our analysis is based on the notion of invariance modulo gauge transformations. This is not only a convenient framework to discuss symmetries in the presence of unphysical degrees of freedom, but it also allows one to reinterpret certain aspects of gauge theories without the need to invoke the conceptually annoying ``breaking of gauge symmetry''.
\end{abstract}

\maketitle

\pagebreak

\section{Introduction}

Physical systems are usually characterized by symmetries that underlie some of their most basic properties. Accordingly, any sound formulation of such systems should make these physical symmetries as explicit as possible.\footnote{The term explicit carries here no prejudice on whether the symmetry is spontaneously broken or not.} Although this should definitely be the case at an exact level, this is not always easy to achieve in the presence of approximations, however.

 The situation is even more intricate in the case where the formulation relies on gauge field theories which contain unphysical degrees of freedom, and, correspondingly, unphysical (gauge) symmetries. Most treatments of these systems require working in a given gauge and, even though the gauge fixing should in principle not interfere with the way physical symmetries constrain the observables, it so happens that, depending on the choice of gauge and depending on the considered symmetry, it is more or less easy to devise approximation schemes that make the associated constraints explicit. Moreover, it is sometimes convenient to work with gauge variant quantities such as the effective action for the gauge field or the corresponding vertex functions. In a chosen gauge, these quantities do not necessarily reflect the physical symmetries of the problem, not even at an exact level. This is certainly an inconvenient feature, in particular within continuum approaches that give primarily access to correlation functions \cite{vonSmekal:1997ohs, Alkofer:2000wg, Zwanziger:2001kw, Fischer:2003rp, Bloch:2003yu, Aguilar:2004sw, Boucaud:2006if, Aguilar:2007ie, Aguilar:2008xm, Boucaud:2008ky, Fischer:2008uz, Rodriguez-Quintero:2010qad,Wetterich:1992yh, Berges:2000ew, Pawlowski:2003hq, Fischer:2004uk,Pawlowski:2005xe,Cyrol:2016tym,Dupuis:2020fhh,Fister:2011uw, Fischer:2012vc,Huber:2012kd, Quandt:2015aaa}.
 
The goal of this work is, first, to identify possible sources for these difficulties and, second, to use this knowledge in order to put forward gauge fixings that are more favorable to the building of symmetry preserving approximations, both at the level of the observables and at the level of the effective action/vertex functions for the gauge field. Although the discussion applies to any type of physical symmetry, we shall first consider the case of center symmetry, which is relevant for the study of the confinement/deconfinement phase transition in pure Yang-Mills (YM) theories at finite temperature. Beyond this particular application, center symmetry provides in fact a prototype for how physical symmetries need to be considered in the presence of unphysical degrees of freedom. 

As for the gauge fixing procedure, we shall consider a rather general formulation which we refer to as {\it gauge fixing on average} and which contains the usual, {\it conditional gauge fixings} as particular limits. Although admittedly a bit formal, it has the advantage of being more rigorous than the Faddeev-Popov approach for it allows one to incorporate possible Gribov copies. Related to this point, we mention that, beyond the question of approximations, it is often necessary to model the gauge fixing procedure in the infrared. This can also strongly interfere with the physical symmetries if the model is constructed within an arbitrary gauge. The gauges that we shall identify are also more favorable to the building of symmetry preserving models of their IR completion.

The paper is organized as follows.  The next two sections introduce the various concepts to be used throughout this work. In Sec.~\ref{sec:center}, we recall basic knowledge about center symmetry as a prototype for how, in general, physical symmetries need to be discussed in the presence of gauge fields. We pay particular attention to the notion of invariance modulo gauge transformations. In Sec.~\ref{sec:gauges}, we introduce the general class of gauge fixings on average upon which we base the subsequent analysis.  After these introductory considerations, and using the particular example of the Polyakov loop, Sec.~\ref{sec:constraints} discusses the physical symmetry constraints on observables and how approximations within a generic gauge might violate these constraints. Section \ref{sec:EA} considers the case of the effective action/vertex functions for the gauge field and shows that, in general, and even at an exact level of treatment, the symmetries do not constrain these objects in a specified gauge but, rather, connect them to the same objects in another gauge. It follows for instance that, unless the gauge is chosen appropriately, the one-point function (obtained from the minimization of the effective action) cannot be used {\it a priori} as an order parameter for the symmetry at hand. In Sec.~\ref{sec:sym_gf}, we define the notion of {\it symmetric gauge fixings} for which the constraints on observables should be more robust to approximations, and for which the above mentioned identities connecting the effective action/vertex functions in two different gauges turn into actual symmetry constraints in that particular gauge. It also discusses under which conditions a given conditional gauge fixing can be considered symmetric. Section \ref{sec:sym_bg} explicitly constructs symmetric gauge fixings as particular realizations of background field gauge fixings. Finally, in Sec.~\ref{sec:self_bg}, we compare this approach to the more familiar one based on self-consistent backgrounds. After some concluding remarks, the Appendices gather some additional material. Appendix A provides a general discussion of physical symmetries in the presence of gauge fields, beyond the particular example of center symmetry. Appendix B discusses other possible ways to construct symmetric gauge fixings, which, however, turn out to be not as promising as the one based on background gauges introduced in Sec.~\ref{sec:sym_bg}. 

The discussion in this article is mostly centered around continuum methods. Its extension to the lattice setting is left for a future work. Every now and then, however, we shall use the lattice to provide a complementary illustration of some of the discussed concepts and ideas. Also, as already mentioned above, our discussion is based on the notion of invariance modulo gauge transformations. Beyond the specific application discussed in this work, it allows one to reinterpret certain aspects of gauge theories without invoking the somewhat annoying notion of ``broken gauge symmetry''. We illustrate these questions in Secs.~\ref{sec:gauge_break}, \ref{sec:Hosotani} and \ref{sec:Higgs}.

\section{Center symmetry}\label{sec:center}

Center symmetry is a symmetry present in YM theories at finite temperature \cite{Gavai:1982er,Gavai:1983av,Celik:1983wz,Svetitsky:1985ye,Pisarski:2002ji,Greensite:2011zz}. It is physical in the sense that it transforms certain observables such as the Polyakov loop, see below. Yet, it acts on gauge fields modulo unphysical (gauge) transformations. Let us see how this comes about using the particular case of SU(N) YM theories. In what follows, it will be convenient to see the gauge field $A_\mu^a$ as an element $\smash{A_\mu\equiv A_\mu^a t^a}$ of the associated Lie algebra.

\subsection{Action on gauge fields}
Center symmetry arises from the observation that, at finite temperature, the transformations 
\beq
A^U_\mu\,\equiv\,UA_\mu U^{-1}+\frac{i}{g}U\partial_\mu U^{-1}\label{eq:AU}
\eeq
that leave the classical action invariant do not all qualify as symmetries under the functional integral. This is because the periodic boundary conditions of the gauge field at finite temperature that restrict the gauge field as
\beq
A_{\mu}(\tau+\beta,\vec{x})=A_{\mu}(\tau,\vec{x})\,,
\eeq
with $\tau$ the Euclidean time and $\smash{\beta=1/T}$ the inverse temperature, impose the following restriction on the SU(N) transformation field $U(\tau,\vec{x})$:
\beq
U(\tau+\beta,\vec{x})=e^{i 2 \pi k/N}U(\tau,\vec{x})\,,
\label{kk}
\eeq
with $\smash{k=0,...,N-1}$, such that the transformed field $A_\mu^U$ remains periodic. The corresponding group of transformations will be denoted $\mathcal{G}$ in what follows. 

It should be emphasized that, only for the subgroup $\smash{\mathcal{G}_0 < \mathcal{G}}$ corresponding to $\smash{k=0}$, do $U(\tau,\vec{x})$ and $A_{\mu}(\tau,\vec{x})$ have the same boundary conditions and that only these transformations should be considered as genuine gauge transformations in the sense of unphysical transformations that do not alter the state of the system. In contrast, any element of the complementary set ${\cal G}-{\cal G}_0\equiv\{U\in {\cal G}\,|\,U\notin{\cal G}_0\}$ should be seen as a physical transformation that changes the state of the system.\footnote{The complementary set is also denoted ${\cal G}\backslash {\cal G}_0$ sometimes. Here, we avoid this notation, not to introduce any confusion with the quotient group ${\cal G}/{\cal G}_0$ to be introduced below.} In particular, it changes the value of the observable
\beq
\ell\equiv \left\langle \Phi[A]\right\rangle\equiv  \frac{1}{N}\left\langle \!{\rm tr}\,{\cal P}\exp\left\{ig\!\int_0^\beta \!\!\! d\tau A_0^a(\tau,\vec{x})\,t^a\right\}\!\right\rangle,\label{eq:pl}
\eeq
known as the {\it Polyakov loop.} That $\ell$ is an observable follows from the fact that it directly connects to the free energy $F$ of a static quark source in a thermal bath of gluons, $\ell\sim e^{-\beta F}$ \cite{Polyakov:1978vu}.

The complementary set ${\cal G}-{\cal G}_0$ is not a very useful concept, however, since it does not possess a natural group structure. To unveil the group structure associated to the physical transformations, we first label all transformations $\smash{U \in \mathcal{G}}$ as $U_k$, with $k$ the integer that determines the phase in the boundary condition \eqref{kk}. This divides $\mathcal{G}$ into $N$ subsets which we denote $\mathcal{U}_k$. The relevance of these subsets is that the change of the Polyakov loop is the same for any transformation $U_k$ belonging to the same subset ${\cal U}_k$. Thus, what really characterizes the symmetry are not the distinct elements of ${\cal G}$ but rather the distinct subsets ${\cal U}_k$, the various transformations $U_k$ belonging to the same subset ${\cal U}_k$ representing the same physical transformation of the state of the system.

In other words, the physical transformations should be identified with the ${\cal U}_k$ and the set of all possible physical transformations is $\{{\cal U}_0, {\cal U}_1,\dots,{\cal U}_{N-1}\}$. In order to define a group structure on this set, we notice that the transformations $U_k$ obey the relation\footnote{This implies in particular that the $U_0$ form a group: the subset $\mathcal{U}_0$ is the aforementioned subgroup $\mathcal{G}_0$. We notice also that $U_k^{-1}\in {\cal U}_{N-k}$.}
\beq
U_k U'_j=U''_{k+j},
\label{JJ}
\eeq
where the labels are defined modulo $N$. We can then define a group law on the set of physical transformations, by setting
\beq
{\cal U}_k\,{\cal U}_j\equiv{\cal U}_{k+j}.\label{LL}
\eeq
This is the actual group of physical center transformations. It is isomorphic to the group of relative integers modulo $N$, $\smash{\mathds{Z}_N\equiv\mathds{Z}/N\mathds{Z}}$, or to the group of the $N^{\rm th}$ roots of unity, $\smash{Z_N\equiv \{e^{i2\pi k/N}\,|\,k=0,\dots,N-1\}}$, or yet to the center of SU(N), defined as the subgroup of elements of SU(N) that commute with any element of SU(N). 

From Eq.~(\ref{JJ}), we also see that two transformations $\smash{U_k,U'_j\in {\cal G}}$ belong to the same subset, and therefore represent the same physical transformation, if and only if they are related by and element $U_0$ of ${\cal G}_0$ as  $\smash{U_k=U_0U'_j}$. This means that the subsets form equivalence classes under the relation 
\beq
U_k\sim U'_j \Leftrightarrow \exists U_0\in{\cal G}_0\,, U_k=U_0U'_j\,,
\eeq 
and that the set of of physical transformations is nothing but the quotient set ${\cal G}/{\cal G}_0$. That it can naturally be endowed a group structure, see Eq.~(\ref{LL}), relates to the fact that ${\cal G}_0$ is a normal subgroup within ${\cal G}$:
\beq
\forall U_0\in {\cal G}_0\,,\,\,\, \forall U\in {\cal G}\,,\,\,\, UU_0U^{-1}\in {\cal G}_0\,.\label{eq:normal}
\eeq 
The identity element of ${\cal G}/{\cal G}_0$ is nothing but the class generated by the identity element of ${\cal G}$ and corresponds thus to ${\cal G}_0$ itself. The associated physical transformation is just the identity transformation that does not change the state of the system, in line with the fact that this class is made of all the genuine gauge transformations. The other classes correspond to the non-trivial center transformations.

In a certain sense, ${\cal G}/{\cal G}_0$ eliminates the gauge redundancy within ${\cal G}$ due to the presence of the subgroup of unphysical transformations ${\cal G}_0$. It should be emphasized that the global gauge transformations (a.k.a. color rotations) belong to ${\cal G}_0$. Then, we are here assuming that they also correspond to redundancies that do not alter the physical state of the system. We do not have a fully rigorous justification for this choice, but rather a collection of implications that make this choice consistent and even useful at an interpretation level. We shall illustrate them at various instances below.

 \begin{figure}[t]
\vglue-12mm
	\centering
	\includegraphics[ width=8.5cm]{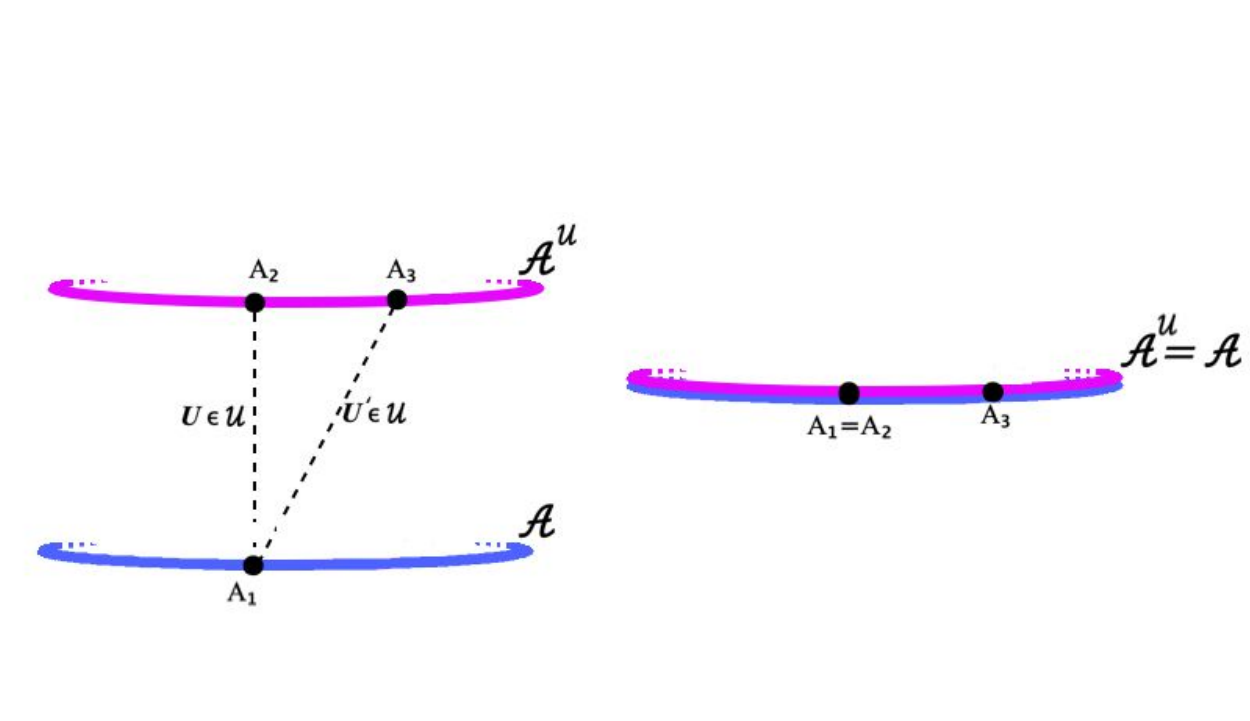}
	\vglue-6mm
	\caption{Schematic representation of two $\mathcal{G}_0$-orbits connected by the gauge class $\mathcal{U}$, with gauge-field configurations $A_1 \in \mathcal{A}$ and $A_2, A_3 \in \mathcal{A}^{\mathcal{U}}$, so that $\smash{A_2^{U_0}=A_3}$.  On the left side, the orbits are not center-symmetric, and $A_1$ is connected to $A_2$ and $A_3$ through elements of $\mathcal{U}$, i.e. $A_1^U=A_2$ and $A_1^{U'}=A_3$. On the right side, the two orbits are center-symmetric and thus lie on top of each other, so that each gauge configuration on $\mathcal{A}$ is identified with one gauge configuration on $\mathcal{A}^{\mathcal{U}}$, in this case $\smash{A_1=A_2=A_1^U}$, this is Eq.~\eqref{eq:4}. Then, two gauge configurations are always connected by a transformation in $\mathcal{G}_0$, in this case $A_1^{U'}=A_3=A_2^{U_0}=A_1^{U_0}$, this is Eq.~\eqref{eq:3}.}\label{fig:df}
\end{figure}

\subsection{States and gauge orbits}\label{sec:classical}

A similar discussion applies to the gauge-field configurations which are redundant due to the possibility of ${\cal G}_0$-transforming them into one another.

At a classical level,\footnote{We shall later adapt this discussion at a quantum level, within a given gauge fixing.} a given gauge-field configuration $A$ can be seen as representing a certain state of the system. However, since ${\cal G}_0$ is the group of gauge transformations, any $A^{U_0}$ with $U_0\in {\cal G}_0$ represents the same physical state. Phrased differently, a physical state is described by a given ${\cal G}_0$-orbit $\smash{{\cal A}=\{A^{U_0}\,|\,U_0\in {\cal G}_0\}}$ and any gauge-field configuration belonging to this orbit is an equivalent representation of that state. By describing the states in terms of ${\cal G}_0$-orbits, one eliminates the gauge redundancy associated to a description in terms of gauge-field configurations, just as ${\cal G}/{\cal G}_0$ removes the gauge redundancy present in ${\cal G}$.

Now, since center transformations are physical transformations, it should be possible to represent them directly on the ${\cal G}_0$-orbits. To see this, we invoke once again the fact that ${\cal G}_0$ is a normal subgroup within ${\cal G}$, Eq.~(\ref{eq:normal}). From this, it is easy to see that the action of ${\cal G}$ on the gauge-field configurations is in fact an action of ${\cal G}/{\cal G}_0$ on the ${\cal G}_0$-orbits: given $U\in {\cal G}$, it transforms all configurations of a given ${\cal G}_0$-orbit into configurations of one and the same ${\cal G}_0$-orbit. Moreover, this ${\cal G}_0$-orbit depends only on the class ${\cal U}$ to which belongs the transformation $U$,\footnote{From now on, we will not display the label $k$ unless a certain value is meant, i.e. $\mathcal{U} \equiv \mathcal{U}_k$ and $U \equiv U_k$ for unspecified $k=0,...,N-1$.} thereby defining the action of ${\cal G}/{\cal G}_0$ on the ${\cal G}_0$-orbits. In what follows, we denote by ${\cal A}^{{\cal U}}$ the transformation of a given ${\cal G}_0$-orbit ${\cal A}$ under ${\cal U}\in {\cal G}/{\cal G}_0$.
 
\onecolumngrid
\begin{center}
\begin{figure}[t]
\includegraphics[height=0.16\textheight]{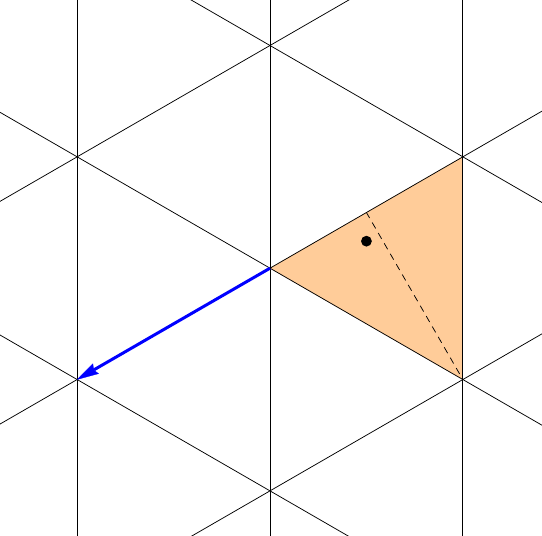}\qquad\includegraphics[height=0.16\textheight]{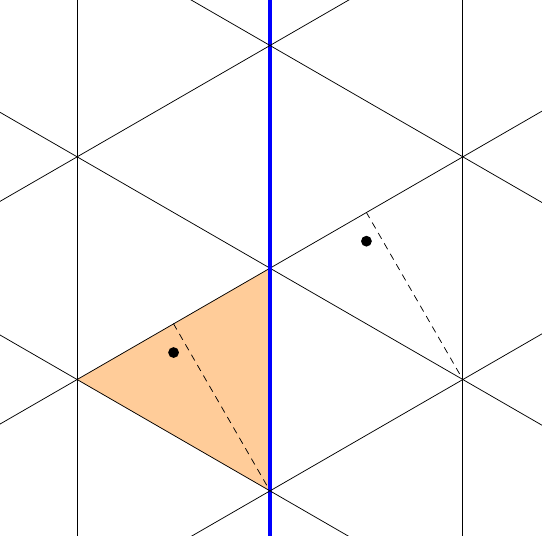}\qquad\includegraphics[height=0.16\textheight]{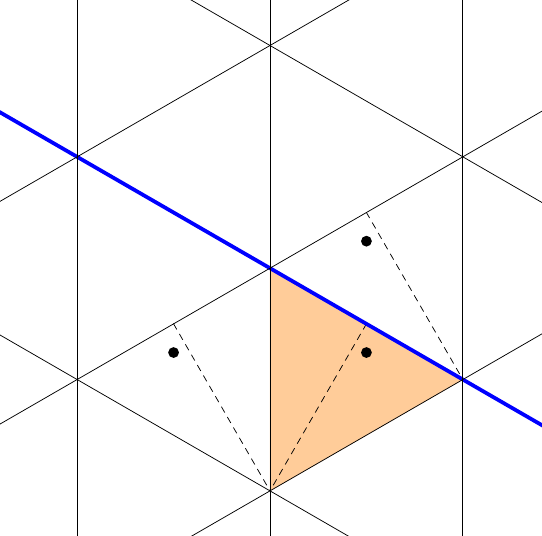}\qquad\includegraphics[height=0.16\textheight]{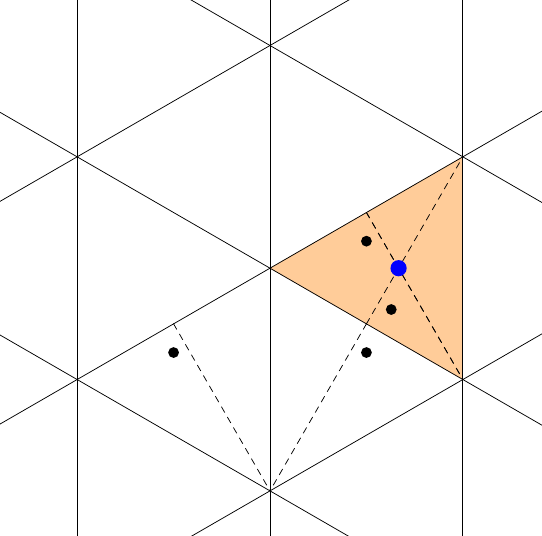}
\caption{Transformation of a Weyl chamber under a center transformation made of a non-periodic element of ${\cal G}$ (corresponding to a translation along one of the edges of the Weyl chamber) followed by two genuine gauge transformations (corresponding to reflections with respect to the edges of the Weyl chambers). The combination of these transformations leads to a specific geometrical transformation of the Weyl chamber into itself. We have chosen a point and a particular axis of the Weyl chamber to ease orientation as the Weyl chamber is transformed. In the first three figures, the blue items represent the transformations that will be applied to the Weyl chamber, while in the fourth figure, the blue item represents the resulting transformation of the original Weyl chamber into itself, here a rotation by an angle $2\pi/3$ around its center.}
\label{fig:SU3_center}
\end{figure}
\end{center}
\twocolumngrid

Of particular interest for the discussion below are those states that are invariant under center symmetry. According to the discussion above, these are represented by ${\cal G}_0$-orbits ${\cal A}_c$ such that
\beq
\forall {\cal U}\in {\cal G}/{\cal G}_0\,,\,\,\, {\cal A}_c^{{\cal U}}={\cal A}_c\,.\label{eq:0}
\eeq
As is easily shown, they correspond to gauge-field configurations $A_c$ such that
\beq
\forall {\cal U}\in {\cal G}/{\cal G}_0\,,\,\,\, \exists U\in {\cal U}\,,\,\,\, A_c^U=A_c\,,\label{eq:4}
\eeq
or, equivalently, such that
\beq
\forall U\in {\cal G}\,,\,\,\, \exists U_0\in {\cal G}_0\,,\,\,\, A_c^U=A_c^{U_0}\,.\label{eq:3}
\eeq
We refer to these configurations as {\it center-symmetric.}\footnote{In fact, it is enough that these conditions apply to the particular class ${\cal U}_1$ of center transformations corresponding to the center element $e^{i2\pi/N}$, since then they are satisfied for any other class in ${\cal G}/{\cal G}_0$.}  According to our definition, see Eq.~(\ref{eq:4}), the center-symmetric configurations are invariant only under certain transformations $U$ in each class $\smash{{\cal U}\in{\cal G}/{\cal G}_0}$. Equivalently, they are invariant under ${\cal G}$ modulo ${\cal G}_0$, see Eq.~(\ref{eq:3}).\footnote{We stress that the transformation $U_0$ that enters this equation depends in general on $U$ and is not necessarily unique.}  A graphical illustration of these conditions is shown in Fig.~\ref{fig:df}.

One could wonder why center-invariant configurations could not be defined by requiring them to be strictly invariant under ${\cal G}$, corresponding to Eq.~(\ref{eq:3}) with $\smash{U_0=\mathds{1}}$. Actually, such configurations do not exist. This is because ${\cal G}_0$, and thus ${\cal G}$, contains transformations that simply translate some of the color components of the gauge field and which have then no fixed point in field space.

\subsection{Weyl chambers}\label{sec:Weyl}
For the discussion below, we will not need to identify all center-invariant configurations but just any particular configuration that obeys Eq.~(\ref{eq:4}) or (\ref{eq:3}). In this respect, it is easy to obtain the center-invariant configurations which are, in addition, constant, temporal and Abelian, that is
\beq
\beta g A_\mu(\tau,\vec{x})=\delta_{\mu0}\,r^jt^j\,,\label{eq:conf}
\eeq
where the $t^j$ span the commuting or diagonal part of the color algebra and the $r^j$ are the components of a vector $\smash{r\in\mathds{R}^{N-1}}$.

First, one introduces the subgroup $\tilde {\cal G}$ of ${\cal G}$ that preserves the particular form (\ref{eq:conf}), and similarly the corresponding subgroup $\tilde {\cal G}_0$ of ${\cal G}_0$. The transformations of $\tilde {\cal G}_0$ can then be shown to subdivide the space $\mathds{R}^{N-1}$ into physically equivalent cells known as {\it Weyl chambers,} each point of a Weyl chamber representing one $\tilde{\cal G}_0$-orbit \cite{Reinosa:2015gxn,Reinosa:2019xqq}. The action of the center symmetry group on the orbits  is represented by transformations of a given Weyl chamber into itself, whose fixed points represent center-symmetric configurations. In order to identify these particular transformations, it is easier to first determine the transformation of the Weyl chamber under an element of $\tilde{\cal G}$. These are typically transformations that take the original Weyl chamber into a different one. In a second step, one uses the transformations in $\tilde{\cal G}_0$ to fold the transformed Weyl chamber back on top of the original Weyl chamber. Doing so, one unveils how the center symmetry group acts on a given Weyl chamber and identifies the center-symmetric configurations of the form (\ref{eq:conf}) \cite{Reinosa:2015gxn,Reinosa:2019xqq}.\footnote{We note that the transformation of $\tilde{\cal G}_0$ that allows one to fold back a Weyl chamber that has been transformed by a certain element $\smash{U\in \tilde{\cal G}}$ is precisely the inverse of a transformation $U_0$ complying with Eq.~(\ref{eq:3}) in the case where $A_c$ belongs to the original Weyl chamber.}

Let us illustrate the previous considerations with some well-known examples. In the SU(2) case, $\smash{r\in\mathds{R}}$ and the action of $\tilde {\cal G}_0$ on configurations of the form (\ref{eq:conf}) is generated by $\smash{r\to -r}$ and $\smash{r\to r+4\pi}$. Equivalently, it is generated by $\smash{r\to 4\pi k-r}$ which correspond to reflection symmetries with respect to the points $2\pi k$. It follows that $\mathds{R}$ is divided into the Weyl chambers $[2\pi k,2\pi(k+1)]$, which are all equivalent to each other from a physical point of view. A non-trivial center transformation (with phase $-1$) is represented by $\smash{r\to r+2\pi}$. It transforms a given Weyl chamber $[2\pi k,2\pi(k+1)]$ into $[2\pi (k+1),2\pi(k+2)]$ which can be folded back into the original Weyl chamber using $\smash{r\to 4\pi (k+1) -r}$. It follows that the corresponding center transformation acts on $[2\pi k,2\pi (k+1)]$ as $\smash{r\to 4\pi (k+1/2) -r}$, that is a reflection with respect to the center $2\pi(k+1/2)$ of the Weyl chamber, which is then the center-symmetric point in this Weyl chamber. In practice, one can restrict to the fundamental Weyl chamber $[0,2\pi]$ whose center-symmetric point is $\pi$. 

In the case of SU(3), the Weyl chambers are equilateral triangles and the non-trivial center transformations are represented by rotations by an angle $\pm 2\pi/3$ about their centers which correspond to the center-symmetric points. This is illustrated in Fig.~\ref{fig:SU3_center} together with the procedure that allows one to identify how the center transformations act on the Weyl chambers. One can restrict to the fundamental Weyl chamber corresponding to $(0,0)$, $(2\pi,2\pi/\sqrt{3})$ and $(2\pi,-2\pi/\sqrt{3})$, with the center-symmetric point located at $(4\pi/3,0)$.

\subsection{Extension to any physical symmetry}
As explained in App.~\ref{sec:sym}, the previous discussion extends to any physical symmetry that operates on gauge fields. Thus, in what follows, ${\cal G}$ denotes any group of physical transformations modulo gauge transformations, the actual group of physical transformations being identified with ${\cal G}/{\cal G}_0$. 

At finite temperature, when restricting to configurations of the form (\ref{eq:conf}), one can again make use of the Weyl chambers. Any physical symmetry corresponds to a certain transformation of a given Weyl chamber into itself whose fixed points are the invariant states for this symmetry. In particular, YM theory is invariant under charge conjugation: $\smash{A_\mu\to -A_\mu^{\rm t}\equiv A_\mu^C}$. In the SU(2) case, this transformation is an element of $G_0$, thus any point of the Weyl chamber is invariant and charge conjugation invariance imposes no constraint. In the SU(3) case, in contrast, charge conjugation is not an element of ${\cal G}_0$ and the action on a given Weyl chamber leaves only certain points invariant. Within the fundamental Weyl chamber, these are the points corresponding to $\smash{r_8=0}$. 

These results are in agreement with physical intuition. In the SU($2$) case for instance, if one evaluates the Polyakov loop functional $\Phi[A]$ associated to a quark, see Eq.~(\ref{eq:pl}), for any configuration of the form (\ref{eq:conf}), one obtains the same result as for the functional $\Phi[A]^*$ which is associated to an antiquark. In the SU($3$) case, in contrast, one only arrives at the same conclusion provided one restricts, within the fundamental Weyl chamber, to $\smash{r_8=0}$. We note that this discussion relies crucially on treating global gauge transformations as redundancies, see the remark above. Had we not made this assumption, we would have wrongly discarded configurations that do actually qualify as charge-conjugation invariant.

Let us finally consider an even simpler example in the vacuum. Take for instance classical electrodynamics and consider the vector potential $\smash{\vec{A}(\vec{x})=(\vec{x}\times \vec{B})/2}$. It corresponds to a constant magnetic field $\vec{B}$ and, thus, to a translationally invariant physical field. Yet, the vector potential is not itself translationally invariant, but translationally invariant modulo a gauge transformation:
\beq
\vec{A}(\vec{x}+a) & = & \vec{A}(\vec{x})+\frac{1}{2}\vec{a}\times\vec{B}\nonumber\\
& = & \vec{A}(\vec{x})+\vec{\nabla}\left(\frac{1}{2}\vec{x}\cdot(\vec{a}\times\vec{B})\right).
\eeq
Moreover, there is no way to gauge transform this configuration into a configuration which is strictly translation invariant because the latter would necessarily correspond to a vanishing magnetic field.

 \subsection{A comment on ``gauge symmetry breaking''}\label{sec:gauge_break}

We insist on the fact that the center-symmetric states correspond to ${\cal G}_0$-orbits that are invariant under the action of ${\cal G}$ (or ${\cal G}/{\cal G}_0$ whose action is the same on the ${\cal G}_0$-orbits). However, the gauge-field configurations that make this orbit are not invariant under the action of ${\cal G}$ but rather under the action of ${\cal G}$ modulo ${\cal G}_0$. The same occurs for the charge conjugation invariant states: they correspond to configurations invariant under $A_\mu\to A_\mu^C$ modulo ${\cal G}_0$ but are not necessarily invariant under $A_\mu\to A_\mu^C$. In general, the invariance under a symmetry group at the level of gauge-field configurations needs always to be understood modulo possible gauge transformations belonging to ${\cal G}_0$.

This leads to an interesting interpretation when the symmetry group under scrutiny is ${\cal G}_0$ itself. Indeed, any gauge-field configuration is invariant under ${\cal G}_0$ modulo ${\cal G}_0$ and thus the state this configuration represents is invariant under gauge transformations. This is obvious since the physical states correspond to the ${\cal G}_0$-orbits which are invariant under ${\cal G}_0$ by construction. Therefore, any gauge-field configuration, representing a certain $\mathcal{G}_0$-orbit, is compatible with gauge symmetry. This interpretation complies with the expectation that there should be no way in which one could conclude to the spontaneous breaking of a local gauge symmetry \cite{Elitzur:1975im}. As already mentioned, we are also assuming that global gauge transformations are redundancies as well, and, as such, we do not expect them to be broken. We shall explain in which sense they are not broken when discussing the Higgs mechanism at the end of the present section as well as in Sec.~\ref{sec:Higgs}.\footnote{If the local gauge transformations that are not broken include transformations that become constant at space-time infinity, then the global gauge transformation cannot be broken either. This is because any global gauge transformation can be seen as the combination of two such local gauge transformations. It is sometimes considered that only the gauge transformations which become the identity at space-time infinity correspond to redundancies \cite{Hertzberg:2018kyi,Hertzberg:2019ffc}. Our point of view in this work is slightly different since it also considers as redundant any gauge transformation that can be obtained from the former via a continuous deformation. This is certainly the case of gauge transformations that become constant at space-time infinity, but certainly not the case of the non-trivial center transformations $\smash{U\in {\cal U}_k}$, with $\smash{k=1,\dots,N-1}$.}

In some approaches to Yang-Mills theories with compactified dimensions, see for instance Ref.~\cite{Pastor-Gutierrez:2022rac}, a different point of view is taken in which gauge-field configurations are classified according to the degree of invariance of the untraced Polyakov loop under elements of ${\cal G}_0$. In this case, it may seem that only certain configurations comply with gauge symmetry while the other break it in various possible forms. We believe that a more sound interpretation is the one that relies on invariance modulo ${\cal G}_0$ and for which any configuration is compatible with gauge symmetry in the extended sense discussed above. In this sense, gauge symmetry can never be broken. What can happen, however, is that the system transitions from a state represented by a ${\cal G}_0$-orbit that contains configurations that are strictly invariant under some of the elements of ${\cal G}_0$, to a state represented by a ${\cal G}_0$-orbit that contains no such configurations. The transition between these two types of states could have observable consequences.

The classification in Ref.~\cite{Pastor-Gutierrez:2022rac} is interesting precisely in this sense and can be given an interpretation in terms of the Weyl chambers. In the SU($3$) case, one can argue that the configurations of type $A$ in Table I of Ref.~\cite{Pastor-Gutierrez:2022rac} correspond to the vertices of the Weyl chambers, while those of type $B$ correspond to the edges of the Weyl chambers and finally configurations of type $C$ correspond to points strictly inside the Weyl chambers including the center-symmetric points. The interesting question is whether the system could transition between any of these configurations. This would not question our interpretation that a gauge symmetry cannot be spontaneously broken, in the sense that any of these configurations is compatible with gauge symmetry as explained above, but this could leave an imprint on some observables. To discuss this question further, we need to leave the classical framework and move to the quantum framework and in particular specify the gauge fixing. We shall go back to this question in Sec.~\ref{sec:self_bg}.

Similar considerations apply to gauge field theories coupled to a Higgs field. Since the latter transforms as $\smash{\varphi\to U_0\varphi}$, one could be tempted to associate a non-zero classical Higgs configuration which minimizes the (classical) Higgs potential $(\varphi^{\dagger}\varphi-v^2)^2$, with the breaking of gauge symmetry. However, any gauge transformation of $\smash{\varphi= v}$,  which also minimizes the potential, is a representative of the same ${\cal G}_0$-orbit, which is invariant under ${\cal G}_0$ by construction. In this sense, any pair $(A,\varphi)$ is compatible with gauge symmetry, irrespectively of  whether $\smash{\varphi=0}$ or $\smash{\varphi\neq 0}$. Again, a transition from $\smash{\varphi=0}$ to $\smash{\varphi\neq 0}$ could leave an imprint on certain observables, but this would not mean that gauge symmetry is broken (since the orbit, as a whole, would remain gauge-invariant) but rather that the system transitions from a state described by an orbit that contains configurations of the Higgs-field that are strictly invariant under the considered transformations to a state described by an orbit that contain no such configurations. We postpone the discussion of the quantum case to Sec.~\ref{sec:self_bg}.

\section{Gauge fixing}\label{sec:gauges}
In order to study how the gauge-fixing procedure might interfere with physical symmetries, let us now introduce a formal gauge-fixing procedure that contains the usual procedures as particular cases.

\subsection{Gauge fixing on average}
Suppose that we can find a functional $\rho[A]$ such that, for any configuration $A$,
\beq
0<\int_{{\cal G}_0} {\cal D}U_0\,\rho[A^{U_0}]<\infty\,.\label{eq:6}
\eeq
In other words, the integral of $\rho[A]$ along any ${\cal G}_0$-orbit should be non-zero and finite.\footnote{We shall assume that such functionals exist although it might be difficult to construct one explicit example.} When this condition is met, we can immediately define another functional
\beq
z[A]\equiv \frac{\rho[A]}{\int_{{\cal G}_0} {\cal D}U_0\,\rho[A^{U_0}]}\,.\label{eq:zofA}
\eeq
It is such that, for any configuration $A$,
\beq
\int_{{\cal G}_0} {\cal D}U_0\,z[A^{U_0}] & = & \int_{{\cal G}_0} {\cal D}U_0\,\frac{\rho[A^{U_0}]}{\int_{{\cal G}_0} {\cal D}U'_0\,\rho[(A^{U_0})^{U_0'}]}\nonumber\\
& = & \int_{{\cal G}_0} {\cal D}U_0\,\frac{\rho[A^{U_0}]}{\int_{{\cal G}_0} {\cal D}U'_0\,\rho[A^{U_0'U_0}]}\nonumber\\
& = & \int_{{\cal G}_0} {\cal D}U_0\,\frac{\rho[A^{U_0}]}{\int_{{\cal G}_0} {\cal D}U'_0\,\rho[A^{U'_0}]}=1\,.\label{eq:8}
\eeq
In the first steps, we have used $\smash{(A^{U_0})^{U'_0}=A^{U_0'U_0}}$, as it can be checked from Eq.~(\ref{eq:AU}), while in the last steps, we have considered the change of variables $U'_0\to U'_0U^{-1}_0$ that changes neither the integration measure\footnote{The Haar measure can be taken right-invariant.} ${\cal D}U'_0$ nor the integration domain. 

From Eq.~(\ref{eq:8}), we see that the functional $z[A]$ provides a {\it partition of unity} along each ${\cal G}_0$-orbit. It can then be used to ``fix the gauge''. To this purpose, one rewrites the expectation value of any ${\cal G}_0$-invariant functional ${\cal O}[A]$ as
\beq
\langle {\cal O}[A]\rangle & = & \frac{\int {\cal D}A\,{\cal O}[A]\,e^{-S_{YM}[A]}}{\int {\cal D}A\,e^{-S_{YM}[A]}}\nonumber\\
& = & \frac{\int {\cal D}A\,\int_{{\cal G}_0} {\cal D}U_0\,z[A^{U_0}]\,{\cal O}[A]\,e^{-S_{YM}[A]}}{\int {\cal D}A\,\int_{{\cal G}_0} {\cal D}U_0\,z[A^{U_0}]\,e^{-S_{YM}[A]}}\nonumber\\
& = & \frac{\int_{{\cal G}_0} {\cal D}U_0\,\int {\cal D}A^{U_0}\,z[A^{U_0}]\,{\cal O}[A^{U_0}]\,e^{-S_{YM}[A^{U_0}]}}{\int_{{\cal G}_0} {\cal D}U_0\,\int {\cal D}A^{U_0}\,z[A^{U_0}]\,e^{-S_{YM}[A^{U_0}]}}\nonumber\\
& = & \frac{\int_{{\cal G}_0} {\cal D}U_0\times\int {\cal D}A\,z[A]\,{\cal O}[A]\,e^{-S_{YM}[A]}}{\int_{{\cal G}_0} {\cal D}U_0\times\int {\cal D}A\,z[A]\,e^{-S_{YM}[A]}}\nonumber\\
& = & \frac{\int {\cal D}A\,z[A]\,{\cal O}[A]\,e^{-S_{YM}[A]}}{\int {\cal D}A\,z[A]\,e^{-S_{YM}[A]}}\,.\label{eq:9}
\eeq
In going from the first to the second line, we have used Eq.~(\ref{eq:8}), while in going from the second to the third line, we have used the invariance under ${\cal G}_0$ of the measure, the observable and the action. Finally, in going from the third to the fourth line, we have considered the change of variables $A\to A^{U_0^{-1}}$. This eventually allows us to factor and cancel the volume of the gauge group: the redundant summation over each orbit in the numerator and the denominator of the first line has been replaced by a non-redundant summation in which each configuration of a given orbit contributes only  partially, that is with a weight $z[A]$, to the numerator and the denominator of the considered observable. 

We refer to this type of gauge fixing as {\it gauge fixing on average.}\footnote{At our formal level of treatment, we are not interested in the practicality of the gauge-fixing procedure, such as the possibility of formulating it as a local renormalizable field theory, the absence of a sign problem, and so on. Of course, these considerations are important in practice.} The reason why the gauge is ``fixed'' is that, contrary to the numerator/denominator in the first line, the contribution of each orbit to the numerator/denominator in the last line is finite.  In particular, one can extend the expression in the last line to gauge-variant functionals. Of course, in the case of observables, the expression in the last line does not depend on $z[A]$ because we can take the steps in Eq.~(\ref{eq:9}) backwards. This relies on the fact that  $z[A]$ is a partition of unity along each ${\cal G}_0$-orbit. We shall return back to this remark below.

\subsection{Conditional gauge fixing}
One particular case of gauge fixing on average is that of {\it conditional gauge fixings} corresponding to the choice 
\beq
\rho[A]=\delta(F[A])\,,\label{eq:dofA}
\eeq
where $F[A]$ is known as the {\it gauge-fixing functional} and $F[A]=0$ as the {\it gauge-fixing condition.} One has in this case\footnote{The notation in the second line of Eq.~(\ref{eq:10}) is somewhat formal. A more explicit form can be obtained by locally considering a chart $U_0(\theta)$ and peforming the integral of $\delta(F[A^{U_0(\theta)}])$ with respect to $\theta$ while taking into account the Haar measure expressed in these variables. The simplest choice is to use a local chart $e^{i\theta^a t^a}U_0^{(i)}(A)$ in the vicinity of each of the $U_0^{(i)}(A)$, with  ${\rm tr}\,\smash{t^a t^b=1}$. In this case, the Haar measure 
\beq
\left.{\rm det}_{ab}\,{\rm tr}\, \frac{\partial e^{-i\theta^ct^c}}{\partial\theta^a}\frac{\partial\partial e^{i\theta^dt^d}}{\partial\theta^b}\right|_{\theta\to 0}={\rm det}_{ab}\,{\rm tr}\,t^a t^b={\rm det}_{ab}\,\delta^{ab}=1\nonumber
\eeq
 contributes trivially and the derivative in the RHS of Eq.~(\ref{eq:10}) needs to be interpreted as $\delta F[(A^{U_0^{(i)}(A)})^{e^{i\theta^at^a}}]/\delta\theta^b|_{\theta\to 0}$.} 
\beq
\int_{{\cal G}_0} {\cal D}U_0\,\rho[A^{U_0}] & = & \int_{{\cal G}_0} {\cal D}U_0\,\delta(F[A^{U_0}])\nonumber\\
& = & \sum_i \left|{\rm det}\,\left.\frac{\delta F[A^{U_0}]}{\delta U_0}\right|_{U_0^{(i)}(A)}\right|^{-1}\,,\label{eq:10}
\eeq
where $\smash{U_0^{(i)}(A)\in{\cal G}_0}$ is a solution to the gauge-fixing condition on the ${\cal G}_0$-orbit of $A$, that is $F[A^{U_0^{(i)}(A)}]=0$. The functional $z[A]$ is then
\beq
z[A] & = & \frac{\delta(F[A])}{\sum_i \left|{\rm det}\,\left.\frac{\delta F[A^{U_0}]}{\delta U_0}\right|_{U_0^{(i)}(A)}\right|^{-1}}\,.\label{eq:11}
\eeq
In order for the right-hand side of Eq.~(\ref{eq:10}) to comply with the left constraint of Eq.~(\ref{eq:6}), we need to assume that each ${\cal G}_0$-orbit is intersected at least once by the gauge-fixing condition. This is a natural assumption since otherwise the gauge-fixing condition could miss configurations that might be important for certain observables. On the other hand, having Gribov copies along each orbit \cite{Gribov77}, that is more than one $U_0^{(i)}(A)$ for each $A$, is not a problem at this formal level of discussion.\footnote{It becomes a problem when trying to rewrite the associated gauge fixing as a tractable functional integral.} We shall only assume that the sum over copies in Eq.~(\ref{eq:10}) is finite, in order to fulfil the right constraint in Eq.~(\ref{eq:6}).

\subsection{Serreau-Tissier gauge fixing}
In the presence of Gribov copies, there exist in fact many other implementations of conditional gauge fixings as gauge fixings on average. Indeed, one can choose $\smash{\rho[A]=w[A]\delta(F[A])}$ provided that
\beq
\sum_i w[A^{U_0^{(i)}(A)}]\left|{\rm det}\,\left.\frac{\delta F[A^{U_0}]}{\delta U_0}\right|_{U_0^{(i)}(A)}\right|^{-1}
\eeq
is non-zero and finite. Noticing that ${\rm det}\,\delta F[A^U]/\delta U$ is only a functional $\Delta [A^U]$ of $A^U$,\footnote{This is because of the interpretation of $\delta F[A^U]/\delta U$ discussed in the previous footnote.} one can even choose $\smash{\rho[A]=w[A]\Delta[A]\delta(F[A])}$ with the condition that
\beq
\sum_i w[A^{U_0^{(i)}(A)}]s[A^{U_0^{(i)}(A)}]
\eeq
is non-zero and finite, with $s[A^{U_0^{(i)}(A)}]$ the sign of the Faddeev-Popov determinant on the Gribov copy $U_0^{(i)}(A)$. This particular choice leads to the class of {\it Serreau-Tissier gauge fixings} \cite{Serreau:2012cg}.\footnote{The original idea \cite{Serreau:2012cg} involved a functional $w[A]$ that allowed for a local formulation. However, this requires the use of the ``replica trick'' which is not void of subtleties.}

\subsection{Case without Gribov copies}
While we remain at a formal level of discussion, we can also consider the case without Gribov copies, that is the case where there is only one $U_0^{(i)}(A)$ for each $A$. In this case, the various implementations of the conditional gauge fixing associated to a given gauge-fixing condition $F[A]$ boil down to\footnote{In the usual Faddeev-Popov gauge fixing, one assumes that this formula is valid even for functionals $F[A]$ presenting Gribov copies. One also removes the absolute value around the Faddeev-Popov determinant. Although these approximations are believed to make sense in the UV, they are not controlled in the IR. In this latter case, it is mandatory to extend the Faddeev-Popov prescription. Possible choices include a partial account of the Gribov copies, as in the Gribov-Zwanziger framework \cite{Gribov77,Zwanziger89,Vandersickel:2012tz,Dudal08}, or a more phenomenological account, as in the Curci-Ferrari framework \cite{Curci:1976bt,Pelaez:2021tpq}.}
\beq
z[A]=\delta(F[A]) \, \left|{\rm det}\left.\frac{\delta F[A^{U_0}]}{\delta U_0}\right|_{U_0=1}\right|.\label{eq:one_copy}
\eeq
We could restrict the functional derivative to $\smash{U_0=1}$ thanks to the presence of the delta function and the assumption that there is only one configuration obeying the gauge-fixing condition on each orbit.

\section{Observables}\label{sec:constraints}

We are now prepared to discuss how the gauge fixing might interfere with physical symmetries. We  consider the case of observables in this section. The next section will be devoted to the effective action for the gauge field and the corresponding vertex functions.

Consider a group of physical symmetries ${\cal G}/{\cal G}_0$ that is realized as ${\cal G}$ at the level of the gauge fields. To these symmetry transformations are associated constraints on certain observables that transform simply enough (most often linearly) under them. The paradigmatic example is that of the Polyakov loop $\ell$
\beq
\ell\equiv \frac{\int {\cal D}A\,\Phi[A]\,e^{-S_{YM}[A]}}{\int {\cal D}A\, e^{-S_{YM}[A]}}
\eeq
associated with center-symmetry transformations. Let us recall how the symmetry constraints emerge in this case, first without gauge fixing and, then, in the presence of  gauge fixing. We shall then argue why, in the latter case, approximations can jeopardize the symmetry constraints.

\subsection{Without gauge fixing}
With $\Phi[A]$ denoting the Polyakov loop functional, see Eq.~(\ref{eq:pl}), we have for any $\smash{U\in {\cal G}}$ 
\beq
\Phi[A^U]=e^{i2\pi k/N}\Phi[A]\,,\label{eq:transfo}
\eeq 
for some $\smash{k=0,\dots,N-1}$ and thus, if the symmetry is not spontaneously broken
\beq
& & \frac{\int {\cal D}A\,\Phi[A]\,e^{-S_{YM}[A]}}{\int {\cal D}A\, e^{-S_{YM}[A]}}\nonumber\\
& & \hspace{0.8cm}=\,\frac{\int {\cal D}A^U\,\Phi[A^U]\,e^{-S_{YM}[A^U]}}{\int {\cal D}A^U\, e^{-S_{YM}[A^U]}}\nonumber\\
& & \hspace{0.8cm}=\,e^{i2\pi k/N}\frac{\int {\cal D}A\,\Phi[A]\,e^{-S_{YM}[A]}}{\int {\cal D}A\, e^{-S_{YM}[A]}}\,,\label{eq:14}
\eeq
where we have used a change of variables $A\to A^U$ and then exploited the invariance of the action\footnote{It is important that the integration domain, here the space of periodic gauge-field configurations, is not changed by the transformation. This is where it becomes important to restrict to transformations obeying Eq.~(\ref{kk}). Transformations that change the boundary conditions will also turn out to be useful below.} and of the integration measure under ${\cal G}$ together with the transformation rule (\ref{eq:transfo}).

One crucial ingredient in these manipulations is that, in the right-hand side of Eq.~(\ref{eq:14}), and up to a phase factor, we have been able to reconstruct the same functional integral as on the left-hand side. Since this integral is nothing but the one that defines the Polyakov loop, we deduce that $\ell=0$ when the symmetry is not broken.\footnote{In the broken symmetry case, one needs to perform these manipulations in the presence of an infinitesimal source coupled to $\Phi[A]$. The symmetry then imposes relations between the Polyakov loop in the presence of various choices of the source \cite{Reinosa:2019xqq}.}

\subsection{With gauge fixing}\label{sec:wgf}
If we repeat the previous steps in the presence of a generic gauge fixing on average, we find instead
\beq
& & \frac{\int {\cal D}A\,\Phi[A]\,z[A]\,e^{-S_{YM}[A]}}{\int {\cal D}A\, z[A]\,e^{-S_{YM}[A]}}\nonumber\\
& & \hspace{0.8cm}=\,\frac{\int {\cal D}A^U\,\Phi[A^U]\,z[A^U]\,e^{-S_{YM}[A^U]}}{\int {\cal D}A^U\, z[A^U]\,e^{-S_{YM}[A^U]}}\nonumber\\
& & \hspace{0.8cm}=\,e^{i2\pi k/N}\frac{\int {\cal D}A\,\Phi[A]\,z[A^U]\,e^{-S_{YM}[A]}}{\int {\cal D}A\, z[A^U]\,e^{-S_{YM}[A]}}\,.\label{eq:15}
\eeq
In order to complete the argumentation as above, we would need to show that the functional integral in the right-hand side is the same as the one in the left-hand side. Of course the simplest scenario would be that where the gauge fixing satisfies $\smash{z[A^U]=z[A]}$. We will come back to this particular requirement below. In general, however, a generic $z[A]$, whose only constraint is to be a partition of unity along ${\cal G}_0$-orbits, has no reason to have simple transformation properties under ${\cal G}$. Thus, in order to connect back to $\ell$, as in Eq.~(\ref{eq:14}), one needs to find another strategy.

We then note that the functional $\smash{z_U[A]\equiv z[A^U]}$ that appears in the right-hand side of Eq.~(\ref{eq:15}) defines a new partition of unity along ${\cal G}_0$-orbits. To verify this we write
\beq
\int_{{\cal G}_0} dU_0\,z_U[A^{U_0}] & = & \int_{{\cal G}_0} dU_0\,z[(A^{U_0})^U]\nonumber\\
& = & \int_{{\cal G}_0} dU_0\,z[A^{UU_0}]\nonumber\\
& = & \int_{{\cal G}_0} dU_0\,z[A^{UU_0U^{-1}U}]\nonumber\\
& = & \int_{{\cal G}_0} dU_0\,z[(A^U)^{UU_0U^{-1}}]\nonumber\\
& = & \int_{{\cal G}_0} dU_0\,z[(A^U)^{U_0}]=1\,.
\eeq
In obtaining $1$ in the last line, we have used Eq.~(\ref{eq:8}) which is valid for any $A$ and thus in particular for $\smash{A\to A^U}$. In the last steps, we have also made the change of variables $\smash{U_0\to U^{-1}U_0U}$ which leaves the integration measure as well as the integration domain invariant.\footnote{Here we assume that the group measure is left- and right-invariant. This is the case for instance when the underlying group is compact. That the integration domain remains invariant relies on the fact that ${\cal G}_0$ is a normal subgroup within ${\cal G}$.} 

Because $z_U[A]$ is a partition of unity, it qualifies as a gauge fixing on average and can be replaced by $z[A]$ in the average appearing in the right-hand side of Eq.~(\ref{eq:15}) since this average is an observable. This allows one to complete the argumentation as in Eq.~(\ref{eq:14}). 

\subsection{Approximations}
The reasoning in the previous section relies on the independence of the expectation value of any observable with respect to the choice of partition of unity. We note, however, that this requires each orbit to be considered entirely. In the presence of approximations, which usually lose contact with the notion of orbit, this property might be invalidated. 

Even though our focus is here on continuum methods, in order to appreciate the problem further, it is useful here to imagine how a gauge fixing on average would be implemented on the lattice where one usually evaluates observables or correlation functions from a certain number of links. In the presence of a gauge fixing on average $z$, the contribution of each link would need to be weighted by the corresponding value of $z$. Now, when computing an observable, because the latter depends only on the gauge orbits, when evaluating all the links along a given orbit, the contribution of $z$ would add up to $1$, and the result of the observable would not depend on $z$. In practice however, only a subset of links would be considered along a given orbit and, therefore, $z$ would not add up to $1$, leaving a residual dependence on $z$ in the evaluation of the observable that would compromise the deduction of symmetry constraints in the presence of approximations. 

Of course, the lattice usually does not rely on gauge fixings on average but rather on conditional gauge fixing, selecting only one configurations per orbit which suffices to ensure the gauge-fixing independence of the observable and thus the symmetry constraints. Unfortunately, continuum conditional gauge fixing is not equivalent to conditional gauge fixing on the lattice. For instance, if Gribov copies are present a rigorous treatment such as the one in Eq.~(\ref{eq:11}) is more similar to a full-fleshed gauge fixing on average in the sense that various (infinitely many) configurations should be selected per orbit in order to ensure the gauge-fixing independence of the observables. So, the problem prevails.\footnote{The ideal case of continuum conditional gauge fixing without Gribov copies is the one that resembles most the conditional gauge fixing on the lattice and one could think that there is no problem in this case. We shall discuss this issue more specifically in Sec.~\ref{sec:odt} and will find that, although the situation is indeed more favorable, it is not totally obvious that the problem can be evaded.}

The above discussion shows that, in the continuum, it could be convenient to find particular realizations of the gauge fixing such that the symmetry under scrutiny is more explicit (that is without relying on full integration along the orbits), making it more resistant to approximations. This will be covered in Sec.~\ref{sec:sym_gf}. Before doing so, let us extend our discussion of the symmetry constraints to the case of the effective action for the gauge-field and the corresponding vertex functions.

\section{Effective action\\ and vertex functions}\label{sec:EA}

So far, we have focused our analysis on observables. Another quantity of interest is the effective action for the gauge field and the related vertex functions. In most continuum approaches, these quantities are simpler to evaluate. It is thus a question whether it is possible to study the symmetries and their breaking in terms of them. 

An important difference with respect to observables is that the very definition of the effective action/vertex functions requires a gauge to be specified. Therefore, these quantities are usually gauge-variant. Moreover, as we will see, within a generic gauge, the symmetries do not impose any constraint on these objects. Rather, they connect them to similar objects in another gauge. This is true even in the absence of approximations.

\subsection{Definitions}
Let us choose a gauge $z[A]$ and define the generating functional for connected correlation functions in that gauge as
\beq
W_z[J]\equiv \ln \int {\cal D}A\,z[A]\,e^{-S_{YM}[A]+J\cdot A}\,,\label{eq:29}
\eeq
where 
\beq
X\cdot Y\equiv \int_x 2\,{\rm tr} X_\mu(x) Y_\mu(x)\,.
\eeq

\vglue1mm

\noindent{The effective action $\Gamma_z[A]$ is then defined as the Legendre transform of $W_z[J]$ with respect to the source $J$:}
\beq
\Gamma_z[A]\equiv -W_z[J_z[A]]+J_z[A]\cdot A\,,
\eeq
with $J_z[A]$ the functional inverse of 
\beq
A_z[J]\equiv \frac{1}{2}\frac{\delta W_z}{\delta J^{\rm t}}\,.
\eeq 
As usual, we have 
\beq
J_z[A]=\frac{1}{2}\frac{\delta\Gamma_z}{\delta A^{\rm t}}\,.
\eeq

\vglue2mm

\noindent{Let us now analyze the consequences of center symmetry on the functionals $W_z[J]$ and $\Gamma_z[A]$.}\\

\subsection{Symmetry identity for $W_z[J]$}
Let us proceed as in Eq.~(\ref{eq:15}). Noting that\footnote{Strictly speaking, this identity applies to the case of center symmetry. For other symmetry groups ${\cal G}$, the identity will be different but we should assume for simplicity that it takes again the form of an affine relation involving $A$.}
\beq
J \cdot A^U=J^{U^{-1}}\!\!\!\cdot A+\frac{i}{g}\,J\cdot(U\partial U^{-1})\,,
\eeq 
with $\smash{J^U\equiv UJU^{-1}}$, we write
\begin{widetext}
\beq
& & \ln \int {\cal D}A\,z[A]\,e^{-S_{YM}[A]+J\cdot A}=\ln \int {\cal D}A^U\,z[A^U]\,e^{-S_{YM}[A^U]+J\cdot A^U}\nonumber\\
& & \hspace{1.0cm}=\,\frac{i}{g} J\cdot(U\partial U^{-1})+\ln \int {\cal D}A\,z[A^U]\,e^{-S_{YM}[A]+J^{U^{-1}}\!\!\!\cdot A}\nonumber\\
& & \hspace{1.0cm}=\,\frac{i}{g}J\cdot(U\partial U^{-1})+\ln \int {\cal D}A\,z_U[A]\,e^{-S_{YM}[A]+J^{U^{-1}}\!\!\!\cdot A}\,.
\eeq
\end{widetext}
We have seen above that $\smash{z_U[A]\equiv z[A^U]}$ is a partition of unity and thus qualifies as a gauge-fixing functional. We thus arrive at
\beq
W_z[J]=W_{z_U}[J^{U^{-1}}]+\frac{i}{g}\,J\cdot(U\partial U^{-1})\,.\label{eq:rel}
\eeq
We stress that $W_z[J]$ depends on $z$, even at an exact level. Therefore, the previous identity is in general not a symmetry constraint on the generating functional within a given gauge, but, rather, a relation between the generating functionals in two different gauges, corresponding to $z$ and $z_U$ respectively. Similarly, by taking $J$-derivatives of Eq.~(\ref{eq:rel}), one deduces relations between the correlation functions in these two gauges but no particular constraints on the correlation functions within a  given gauge.

\vglue-10mm

\subsection{Symmetry identity for $\Gamma_z[A]$}
The previous considerations translate into similar results for the effective action $\Gamma_z[A]$. First, from Eq.~(\ref{eq:rel}), one deduces that
\beq
A_z[J]=UA_{z_U}[J^{U^{-1}}]U^{-1}+\frac{i}{g}U\partial U^{-1}=A^U_{z_U}[]\,,\nonumber\\
\eeq
which, upon functional inversion, leads to\footnote{Indeed, $A_z[J^U_{z_U}[A^{U^{-1}}]]=A^U_{z_U}[J_{z_U}[A^{U^{-1}}]]=(A^{U^{-1}})^U=A$.}
\beq
J_z[A]=J^U_{z_U}[A^{U^{-1}}]\,,
\eeq
and thus
\begin{widetext}
\beq
\Gamma_z[A] & = & -W_z[J_z[A]]+J_z[A]\cdot A\nonumber\\
& = & -W_{z_U}[J_{z_U}[A^{U^{-1}}]]-\frac{i}{g}\,J^U_{z_U}[A^{U^{-1}}]\cdot(U\partial U^{-1})+J^U_{z_U}[A^{U^{-1}}]\cdot A\nonumber\\
& = & -W_{z_U}[J_{z_U}[A^{U^{-1}}]]+J_{z_U}[A^{U^{-1}}]\cdot A^{U^{-1}}\,,
\eeq
\end{widetext}
that is
\beq
\Gamma_z[A]=\Gamma_{z_U}[A^{U^{-1}}]\,.\label{eq:rel2}
\eeq
Once again, because $\Gamma_z[A]$ depends on $z$, this is in general not a symmetry constraint within a given gauge but, rather, a relation between the effective actions in two different gauges corresponding to $z$ and $z_U$ respectively. From this identity, one deduces relations between the vertex functions in these two gauges but, a priori, no particular constraints on the vertex functions within a given gauge.

Another consequence of the above is that the minimum $A_{\rm min}[z]$ of $\Gamma_z[A]$ does not qualify in general as an order parameter for the symmetry. Indeed, the absence of a symmetry constraint on $\Gamma_z[A]$ prevents one from deducing that the minimum is invariant under the symmetry in the case where the minimum is unique (contrary to what happens for an actual order parameter when the symmetry is realized in the Wigner-Weyl sense).

All these inconvenient features will find a solution in Sec.~\ref{sec:sym_gf} once we introduce the notion of symmetric gauge fixings.

\subsection{Free energy}
We stress that, despite the functional $W_z[J]$ being $z$-dependent, its zero-source limit $W_z[0]$ should be $z$-independent as it corresponds to a physical quantity: the free-energy of the system. Correspondingly, the value of $\Gamma_z[A]$ at its absolute minimum $A_{\rm min}[z]$ should be independent of $z$:\footnote{That the limit of zero sources corresponds to an extremum of $\Gamma_z[A]$ follows from the fact that $\smash{\delta\Gamma_z[A]/\delta A=2J^{\rm t}}$. That it corresponds to an absolute minimum follows from the convexity of $W_z[J]$ which should be guaranteed if $z[A]$ is not too negative.}
\beq
0=\frac{\delta}{\delta z}\Gamma_z[A_{\rm min}[z]]\,.\label{eq:z_ind}
\eeq
This is just the statement that the same physics should be accessible from any gauge. 

Then, at an exact level, the fact that the symmetry is not explicit on the effective action $\Gamma_z[A]$ for a given, generic $z$, is not really a problem. Indeed, it is enough that it is explicit for one particular choice of $z$. Then, if the symmetry breaks in that gauge at the level of the effective action, this will leave an imprint on (the derivatives of) the free energy, which will then be carried out to any other gauge through the identity (\ref{eq:z_ind}). This explains in particular how the deconfinement transition can occur, at least in principle, in gauges that preserve color rotation invariance and for which $\smash{A_{\rm min}[z]=0}$ at any temperature. 

Unfortunately, the identity (\ref{eq:z_ind}) is difficult to maintain within an approximation set-up.\footnote{In a strict perturbative expansion, in a scheme that does not depend on $z$, Eq.~(\ref{eq:z_ind}) is true order by order. However, it is very often the case that perturbation theory is not applied to $\Gamma_z[A_{\rm min}[z]]$ but rather to $\Gamma_z[A]$ in which case $\Gamma_z[A_{\rm min}[z]]$ contains all orders, some of them only partially, and then Eq.~(\ref{eq:z_ind}) applies only approximately. Beyond perturbation theory, for instance in the context of Dyson-Schwinger equations or within the functional Renormalization Group framework, it is even more true that Eq.~(\ref{eq:z_ind}) is not exactly fulfilled by most truncations.} This is yet another reason for looking for gauges where the symmetry is explicit at the level of the effective action.

\section{Symmetric gauge fixings}\label{sec:sym_gf}
We now define a class of gauge fixings that make the symmetry explicit and that solve the issues described in the previous two sections. We first define this particular class of gauge fixings within the very generic class of gauge fixings on average and then specify to the case of conditional gauge fixings.

\subsection{Gauge fixings on average}
In Sec.~\ref{sec:wgf}, we have already mentioned that, within a gauge-fixed context, one way to retrieve the symmetry constraint on the Polyakov loop would be to find a functional $z[A]$ that is explicitly invariant under ${\cal G}$
\beq
\forall U \in {\cal G}\,, \quad z[A^U]=z[A]\,.\label{eq:17}
\eeq
However, we can immediately argue that there is no such functional. Indeed, if there were, Eq.~(\ref{eq:17}) would be valid for any $\smash{U\in {\cal G}_0}$ and this would immediately imply that $z[A]$ integrates to infinity along each orbit, invalidating the assumption that it is a partition of unity.

One could argue that the condition (\ref{eq:17}) is too strong since it includes that part of the symmetry that has to do with unphysical (gauge) transformations and that the good requirement is rather
\beq
\forall U \in {\cal G}-{\cal G}_0\,, \quad z[A^U]=z[A]\,.\label{eq:18}
\eeq
But there again, it is easy to show that elements of ${\cal G}-{\cal G}_0$ generate ${\cal G}_0$, so that Eq.~(\ref{eq:18}) implies Eq.~(\ref{eq:17}) and one ends up in the same dead end as before.

In fact, the requirement (\ref{eq:18}) is still too strong. However, we only need to require that $z[A]$ is invariant under ${\cal G}$ modulo ${\cal G}_0$, that is
\beq
\forall {\cal U}\in {\cal G}/{\cal G}_0\,,\,\,\, \exists U\in {\cal U}\,,\,\,\, z[A^U]=z[A]\,.\label{eq:20}
\eeq
Contrary to the previous conditions (\ref{eq:17}) or (\ref{eq:18}), there is no obvious argument to discard this possibility. In Sec.~\ref{sec:sym_bg}, we shall actually construct explicit realizations of such gauge fixings using background-based gauge fixings. 

For each $U$ appearing in Eq.~(\ref{eq:20}), it is now possible to perform a similar argumentation as in Eq.~(\ref{eq:14}). One writes
\beq
& & \frac{\int {\cal D}A\,z[A]\,\Phi[A]\,e^{-S_{YM}[A]}}{\int {\cal D}A\, e^{-S_{YM}[A]}}\nonumber\\
& & \hspace{0.8cm}=\,\frac{\int {\cal D}A^U\,z[A^U]\,\Phi[A^U]\,e^{-S_{YM}[A^U]}}{\int {\cal D}A^U\, e^{-S_{YM}[A^U]}}\nonumber\\
& & \hspace{0.8cm}=\,e^{i2\pi k/N}\frac{\int {\cal D}A\,z[A]\,\Phi[A]\,e^{-S_{YM}[A]}}{\int {\cal D}A\, e^{-S_{YM}[A]}}\,.
\eeq
The important point is that we have now been able to find a constraint for the Polyakov loop in the gauge $z[A]$ without ever invoking the partition of unity nature of $z[A]$ and therefore the notion of orbit. The constraint for the Polyakov loop in this gauge should then be more resistant to approximations. 

Let us also note that the fact that we were only able to use certain transformations $U$ in each of the classes $\smash{{\cal U}\in{\cal G}/{\cal G}_0}$ is not a limiting factor since all the classes of ${\cal G}/{\cal G}_0$ are represented, allowing us to unveil all the consequences of the physical symmetry on the corresponding observables. In what follows, we denote gauge-fixing functionals obeying Eq.~(\ref{eq:20}) as $z_c[A]$ and we refer to them as {\it symmetric gauge fixings.} 

It is also important to stress that, within the particular example of center symmetry, we are, in a sense, following a ``top-down approach'', that is, starting from a very large group of symmetries ${\cal G}$ containing redundancies related to the subgroup ${\cal G}_0$, we remove these redundancies by considering the physical group of symmetries ${\cal G}/{\cal G}_0$. As we have argued, it is clear that a given gauge fixing functional $z[A]$ cannot be invariant under the whole ${\cal G}$, but it could be invariant under certain transformations in each of the classes of ${\cal G}/{\cal G}_0$. In some other cases, instead, one starts already from a physical group of symmetries ${\cal T}$, free of any redundancies, and it can happen that $z[A]$ is invariant under ${\cal T}$ from the start, take for instance the case of translation invariance in the Landau gauge. In the case where it is not invariant, it could still happen that, in a sort of ``bottom-up approach", after enlarging ${\cal T}$ into a group $\smash{{\cal G}\equiv {\cal T}\ltimes {\cal G}_0}$ that contains the gauge transformations, the gauge-fixing functional $z[A]$ is found to be invariant under one transformation in each of the classes of ${\cal G}/{\cal G}_0$, transformations which are not necessarily the original transformations within ${\cal T}$, see App.~\ref{sec:sym} for more details.

\subsection{Properties}
Symmetric gauge fixings are also particular with regard to the effective action/vertex functions. Indeed, for a symmetric gauge fixing, the identity (\ref{eq:rel}) relating two gauges becomes a symmetry constraint within one single gauge:
\beq
& & \forall {\cal U}\in {\cal G}/{\cal G}_0\,,\,\,\, \exists U\in {\cal U}\,,\,\,\,\nonumber\\
& & \hspace{0.8cm} W_{z_c}[J]=W_{z_c}[J^{U^{-1}}]+\frac{i}{g}\,J\cdot(U\partial U^{-1})\,,\label{eq:symW}
\eeq
from which one can deduce corresponding constraints on the correlation functions in the gauge $z_c$. Similarly, the identity (\ref{eq:rel2}) becomes
\beq
\forall {\cal U}\in {\cal G}/{\cal G}_0\,,\,\,\, \exists U\in {\cal U}\,,\,\,\,\Gamma_{z_c}[A^U]=\Gamma_{z_c}[A]\,,\label{eq:symGamma}
\eeq
from which one can deduce corresponding constraints on the vertex functions in the gauge $z_c$. These constraints provide as many potential order parameters for the symmetry, see Ref.~\cite{in_prep} for a thorough discussion in the SU(2) and SU(3) cases.

In this case also, the minimum $A_{\rm min}[z_c]$ qualifies as an order parameter for the symmetry. Indeed, if the minimum is unique, it needs to satisfy $\smash{A^U_{\rm min}[z_c]=A_{\rm min}[z_c]}$ and, thus, it can only take specific configurations as given by the fixed points of these transformations.  Any deviation of the minimum from these fixed points would signal a breaking of the symmetry.\footnote{Note that if the fixed points exist, they have to be configurations among the center-symmetric configurations introduced in Sec.~\ref{sec:classical}.}

Let us stress that working within a gauge where the effective action is center-symmetric does not necessarily imply that the confinement/deconfinement transition will materialize as the deviation of the minimum from its center-symmetric configuration, see the discussion in App.~C. On the other hand, the symmetric gauge fixings based on symmetric background configurations which we introduce in Sect.~\ref{sec:sym_bg} seem to allow one to observe such deviations, and, therefore to probe the confinement/deconfinement transition. At least, this has been tested within the context of background Landau gauges and their infrared phenomenological completion by means of the Curci-Ferrari model \cite{Reinosa:2014ooa,Reinosa:2014zta,Reinosa:2015gxn}.

\section{Symmetric backgrounds}\label{sec:sym_bg}

So far, we have identified the condition (\ref{eq:20}) for a gauge fixing to be symmetric.\footnote{In the case of conditional gauge fixings based on gauge-fixing conditions $F[A]=0$, it is interesting to analyze under which conditions the associated gauge-fixing functional $z[A]$ is symmetric. We discuss this question in App.~\ref{sec:cond}.} However, we have not yet constructed any explicit example of such gauge fixings. We now put forward a particular realization using the family of background field gauges and the notion of symmetric backgrounds. For other possible strategies, see App.~C. As we did above, we focus on the example of center-symmetry but our discussion can easily be extended to other symmetries.

\subsection{Background field gauges}
In general, background fields allow one to extend certain gauge fixings characterized by a functional $z[A]$ into classes of gauge fixings characterized by a family of functionals $z_{\bar A}[A]$ such that $\smash{z_{\bar A=0}[A]=z[A]}$ and
\beq
\forall U\in {\cal G}\,, \,\,\, z_{\bar A^U}[A^U]=z_{\bar A}[A]\,.\label{eq:30}
\eeq
The background $\bar A$ should be interpreted as an infinite collection of gauge-fixing parameters whose specification selects one particular gauge within the class of gauges. For these gauge fixings, it follows from Eq.~(\ref{eq:rel}) that
\beq
\forall U\in {\cal G}\,, \,\,\, W_{\bar A^U}[J^U]=W_{\bar A}[J]-\frac{i}{g}\,J\cdot U^{-1}\partial U\,,\label{eq:abc}
\eeq
and, correspondingly,
\beq
\forall U\in {\cal G}\,, \,\,\, \Gamma_{\bar A^U}[A^U]=\Gamma_{\bar A}[A]\,.\label{eq:cov}
\eeq
where we have defined $\smash{W_{\bar A}\equiv W_{z_{\bar A}}}$ and $\smash{\Gamma_{\bar A}\equiv\Gamma_{z_{\bar A}}}$ for simplicity. 

These identities should not be mistaken with Eqs.~(\ref{eq:symW}) and (\ref{eq:symGamma}) however. First, they apply to any transformation $U\in{\cal G}$ rather than to certain transformations $U$ in each class $\smash{{\cal U}\in{\cal G}/{\cal G}_0}$. Second, rather than being identities within a specific gauge, they connect two different gauges, corresponding to $\bar A$ and $\bar A^U$. In particular, Eq.~(\ref{eq:30}) cannot be used to deduce symmetry constraints on observables. Moreover, $A_{\rm min}[\bar A]$ does not qualify in general as an order parameter for the symmetry and the corresponding vertex functions in the gauge associated to $\bar A$ are not constrained by the symmetry. Instead, they are related to the corresponding vertex functions in the gauge associated to $\bar A^U$. 

As we now discuss, interestingly enough, there exist particular choices of the background $\bar A$ for which the gauge fixing becomes explicitly symmetric and the above identities turn into symmetry constraints within that gauge.

\begin{figure}[t]
\begin{center}
\includegraphics[height=0.25\textheight]{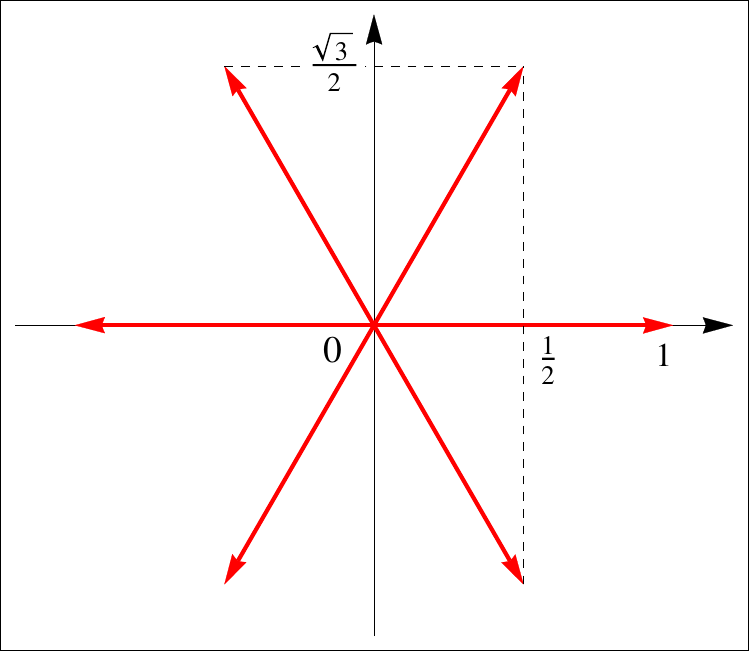}
\caption{Root diagram of the su($3$) algebra.}
\label{fig:roots}
\end{center}
\end{figure}

\subsection{Center-symmetric backgrounds}
Mimicking a discussion that we had above, we could start inquiring about the existence of backgrounds that are invariant under ${\cal G}$. No such backgrounds exists, basically because the group ${\cal G}$ is too large. This is consistent with the fact that this would lead to functionals $z[A]$ that are invariant under ${\cal G}$ and which we have already argued not to exist. Similarly, no choice of background is invariant under ${\cal G}-{\cal G}_0$.

But we could look for backgrounds that are invariant under ${\cal G}$ modulo ${\cal G}_0$. These are precisely the configurations defined in Eqs.~(\ref{eq:4})-(\ref{eq:3}) and for which we have given explicit examples in Sec.~\ref{sec:Weyl}. We shall refer to these choices as {\it center-symmetric backgrounds} and denote them as $\smash{\bar A=\bar A_c}$. It is immediate to check using Eq.~(\ref{eq:30}) that $z_{\bar A_c}[A]$ obeys  Eq.~(\ref{eq:20}). Therefore, it allows one to deduce symmetry constraints on observables without relying on the partition of unity nature of $z_{\bar A_c}[A]$ and it is more prone to allow for the building of approximations or models that keep the symmetry explicit. 

Similarly, from Eqs.~(\ref{eq:4}), (\ref{eq:abc}) and (\ref{eq:cov}), we find that $\forall {\cal U}\in {\cal G}/{\cal G}_0$, $\exists U\in {\cal U}$ such that
\beq
W_{\bar A_c}[J^U]=W_{\bar A_c}[J]-\frac{i}{g}\,J\cdot U^{-1}\partial U\,,
\eeq
and
\beq
\Gamma_{\bar A_c}[A^U]=\Gamma_{\bar A_c}[A]\,.\label{eq:33}
\eeq
Thus, the effective action $\Gamma_{\bar A_c}[A]$ reflects the center symmetry explicitly. In particular, $A_{\rm min}[\bar A_c]$ qualifies as an order parameter for the breaking of the symmetry. Similarly, one expects the vertex functions derived from this effective action to reflect center-symmetry, and, therefore, to allow one to test its breaking. Some of these signatures have been already identified in Ref.~\cite{vanEgmond:2021jyx,vanEgmond:2022nuo} for the SU(2) and SU(3) cases, and will be more thoroughly discussed in a future work \cite{in_prep}. Of notable importance is the fact that the zero-momentum SU($2$) propagator in the center-symmetric Landau gauge diverges at the transition \cite{vanEgmond:2021jyx}.

\subsection{Constant background fields}
We stress that any choice of center-symmetric background is equally good. In particular, we may consider center-symmetric backgrounds of the form
\beq
\beta g \bar A_\mu(\tau,\vec{x})=\delta_{\mu0}\,\bar r^jt^j\,,\label{eq:conf3}
\eeq
with the $\bar r^j$ taking center-symmetric values which we discussed in Sec.~\ref{sec:Weyl} but which we leave unspecified in what follows. One benefit of this choice is that there is an additional symmetry that can be exploited. More precisely, color rotations of the type $\smash{U_\theta\equiv\exp\{i\theta^jt^j\}\in {\cal G}_0}$ leave the background (\ref{eq:conf3}) invariant. Then, Eq.~(\ref{eq:cov}) turns into the symmetry constraint
\beq
\forall \theta\,, \,\,\, \Gamma_{\bar A}[A^{U_\theta}]=\Gamma_{\bar A}[A]\,.\label{eq:theta}
\eeq
The numerous consequences of this symmetry have been discussed in Ref.~\cite{vanEgmond:2022nuo}. In particular, this is a continuous symmetry with associated Noether charges, which, for fields in the adjoint representation, these are the adjoint charges $[t^j,\,\,]$. 

To make this symmetry and the conservation rules explicit, it is convenient to decompose all the adjoint fields along a basis $t^\kappa$ that diagonalizes all these charges simultaneously. This is known as a {\it Cartan-Weyl basis}:
\beq
[t^j,t^\kappa]=\kappa^j t^\kappa\,.
\eeq 
The labels $\kappa$ are to be seen as real-valued vectors in a space isomorphic to the commuting subalgebra. They are of two types. First, any $t^j$ qualifies itself as a $t^\kappa$ with $\smash{\kappa=0}$. To distinguish between the various $t^j$, we shall write instead $\smash{\kappa=0^{(j)}}$, which we refer to as {\it zeros}. This notational trick is only needed when $\kappa$ refers to the label of a generator.\footnote{In equations where it does not play the role of a label, $\kappa$ is a mere vector and, therefore, $0^{(j)}$ can be replaced by $0$. This should always be clear from the context.} The other possible values of $\kappa$ are (non-degenerate) non-zero vectors known as {\it roots} and denoted $\alpha$, $\beta$, \dots The set of roots characterizes the algebra under consideration. In the SU(2) case for instance, there is one zero and two (one-dimensional) roots, $\pm 1$. In the SU(3) case, there are two zeros and six (two-dimensional) roots, $\pm(1,0)$, $\pm(1/2,\sqrt{3}/2)$ and $\pm(1/2,-\sqrt{3}/2)$, see Fig.~\ref{fig:roots}. In what follows, we shall also refer to the zeros and the roots as the {\it neutral modes} and the {\it charged modes,} respectively.

\subsection{Connection to gauges without background}\label{eq:no_bg}
Restricting to backgrounds of the form (\ref{eq:conf3}) allows one to unveil an interesting relation with the gauge $\smash{z_{\bar A=0}[A]=z[A]}$ in the absence of background, which we now describe.

So far, we have restricted to transformations $\smash{U\in {\cal G}}$ that preserve the periodicity of the gauge-field configurations. Very often, however, the relation (\ref{eq:30}) applies to more general transformations that modify the boundary conditions. In particular, for configurations of the form (\ref{eq:conf3}), it is interesting to consider thetransformation
\beq
U_{\bar r}(\tau)\equiv e^{-i\frac{\tau}{\beta}\bar r^jt^j}\,.
\eeq
It is such that
\beq
\beta g\bar A^{U_{\bar r}}_\mu & = & U_{\bar r}\beta g\bar A_\mu U_{\bar r}^{-1}+i\beta U_{\bar r}\partial_\mu U_{\bar r}^{-1}\nonumber\\
& = & \delta_{\mu0}\,U_{\bar r}\bar r^jt^jU_{\bar r}^{-1}-\delta_{\mu0}\,U_{\bar r}\bar r^jt^jU_{\bar r}^{-1}=0\,.
\eeq
It follows that
\beq
z_{\bar A}[A]=z_{\bar A^{U_{\bar r}}}[A^{U_{\bar r}}]=z_0[A^{U_{\bar r}}]\,.
\eeq
This relates the background gauge fixing $z_{\bar A}[A]$ over the space of periodic gauge-field configurations to the corresponding gauge fixing in the absence of background but over the space of gauge-field configurations that are the $U_{\bar r}$-transformed of periodic configurations. We refer to these configurations as {\it background-twisted configurations.} 

To unveil the nature of the background-twisted configurations, let us consider an adjoint periodic field which we decompose along a Cartan-Weyl basis, $\smash{\varphi(\tau)=\varphi^\kappa(\tau) t^\kappa}$, with $\smash{\varphi_\kappa(\tau+\beta)=\varphi_\kappa(\tau)}$. Under the action of $U_{\bar r}$, we have\footnote{We could also include the possibility of an affine term $i/g U_{\bar r}\partial U_{\bar r}^{-1}$ in the transformation of the field but this would not change our conclusion below regarding the boundary conditions of the transformed field.}
\beq
\varphi_{\bar r}(\tau) & \equiv & U_{\bar r} \varphi(\tau) U_{\bar r}^{-1}=e^{-i\frac{\tau}{\beta}\bar r^j[t^j,\,\,]}\varphi(\tau)\nonumber\\
& = & \varphi^\kappa(\tau)e^{-i\frac{\tau}{\beta}\bar r^j[t^j,\,\,]}t^\kappa=e^{-i\frac{\tau}{\beta}\bar r^j\kappa^j}\varphi^\kappa(\tau)t^\kappa\,,\nonumber\\
\eeq
so that $\varphi_{\bar r}^\kappa(\tau)=e^{-i\frac{\tau}{\beta}\bar r^j\kappa^j}\varphi^\kappa(\tau)$. Then,
\beq
\varphi_{\bar r}^\kappa(\tau+\beta) & = & e^{-i\frac{\tau+\beta}{\beta}\bar r^j\kappa^j}\varphi^\kappa(\tau+\beta)\nonumber\\
& = & e^{-i\bar r^j\kappa^j}e^{-i\frac{\tau}{\beta}\bar r^j\kappa^j}\varphi^\kappa(\tau)=e^{-i\bar r^j\kappa^j}\varphi_{\bar r}^\kappa(\tau)\,.\nonumber\\
\eeq
It follows that the background-twisted configurations are periodic modulo a phase factor $e^{-i\bar r^j\kappa^j}$. We have thus found that the neutral modes remain periodic while the charge modes are periodic modulo a phase factor $e^{-i\bar r^j\alpha^j}$.

One can now follow the same steps that lead to Eq.~(\ref{eq:abc}), the only difference being that the integration domain under the functional integral is changed under the change of variables $A\to A^{U_{\bar r}}$. We then find
\beq
W^{\rm twisted}_{\bar A=0}[J^U]=W^{\rm periodic}_{\bar A}[J]-\frac{J_0}{\beta g} \bar r^j t^j\,.
\eeq
In particular, the one-point functions are related by
\beq
A^{\rm twisted}_{\rm min,\mu}[\bar A=0]=A_{\rm min,\mu}[\bar A]-\frac{\delta_{\mu 0}}{\beta g}\bar r^j t^j\,.\label{eq:rel11}
\eeq
Similar relations apply to correlation functions and also to vertex functions. Of particular interest for us is the choice of a center-symmetric background $\smash{\bar r=\bar r_c}$. In this case, Eq.~(\ref{eq:rel11}) rewrites
\beq
A^{\rm twisted}_{\rm min,\mu}[\bar A=0]=A_{\rm min,\mu}[\bar A_c]-\frac{\delta_{\mu 0}}{\beta g}\bar r^j_c t^j\,.
\eeq
Since $A_{\rm min,\mu}[\bar A_c]$ is an order parameter, we conclude that $A^{\rm twisted}_{\rm min,\mu}[\bar A=0]$ is also an order parameter, computed in a gauge without background but in the presence of twisted boundary conditions.

The previous considerations are very useful in regard to a possible implementation of symmetric gauges $\smash{\bar A=\bar A_c}$ on the lattice. Basically, any implementation of a no-background gauge could in principle be adapted into a symmetric gauge by just changing the periodic boundary conditions into twisted boundary conditions. Work in this direction is currently in progress and could potentially bring some interesting twist to state-of-the-art calculations \cite{Cucchieri:2010lgu,Fischer:2010fx,Maas:2011ez,Mendes:2015jea,Silva:2013maa,Aouane:2011fv,Silva:2016onh}.

\section{Self-consistent backgrounds}\label{sec:self_bg}

Before concluding, let us compare the strategy discussed in Sec.~\ref{sec:sym_bg}, based on the use of center-symmetric backgrounds and the minimization of the corresponding center-symmetric effective action $\Gamma_{\bar A_c}[A]$, to the usual strategy based on the minimization of the background effective action (whose definition we recall below) and the associated self-consistent backgrounds.

\subsection{Background field effective action}
Recall that the problem with generic backgrounds (or more generally with generic gauge fixings) is that Eq.~(\ref{eq:cov}) is not a symmetry constraint on the effective action in that gauge. In particular, the corresponding minima are not order parameters. To cope with this, one frequently introduces the functional \cite{Abbott:1980hw,Abbott:1981ke,Braun:2007bx,Braun:2010cy,Reinosa:2015gxn,Reinosa:2019xqq}
\beq
\tilde\Gamma[\bar A]\equiv \Gamma_{\bar A}[A=\bar A]\,,\label{eq:def}
\eeq 
known as the {\it background field effective action.} It is immediate to show that
\beq
\forall U\in {\cal G}\,, \quad \tilde\Gamma[\bar A^U]=\tilde\Gamma[\bar A]\,.\label{eq:bg_inv}
\eeq
Unlike Eq.~(\ref{eq:cov}) which relates two functionals, Eq.~(\ref{eq:bg_inv}) is a symmetry identity on a unique functional, $\tilde\Gamma[\bar A]$. The minima of this functional could then play the role of order parameters, just as the minima of $\Gamma_{\bar A_c}[A]$. There are, however, a few subtleties that need to be discussed before qualifying the minima of  $\tilde\Gamma[\bar A]$ as order parameters. These subtleties relate to the fact that, unlike $\Gamma_{\bar A_c}[A]$, the functional $\tilde\Gamma[\bar A]$ is not a Legendre transform.

First of all, no matter what phase the system is in, $\tilde\Gamma[\bar A]$ always admits infinitely many minima. This is because, as a consequence of Eq.~(\ref{eq:bg_inv}), $\tilde\Gamma[\bar A]$ is invariant under ${\cal G}_0$. This obscures the discussion of physical symmetry breaking which usually corresponds to the transition between a Wigner-Weyl phase with one minimum (whose value is constrained by the symmetry) to a Nambu-Goldstone phase with degenerate minima (connected to each other by the symmetry). The problem has a simple fix, however, since the ${\cal G}_0$-multiplicity of minima is nothing but the expression of gauge redundancy. The latter can be gauged away by defining a background effective action directly on the background orbits $\bar {\cal A}$ rather than on the background configurations $\bar A$ themselves. More precisely, one defines $\smash{\tilde\Gamma[\bar {\cal A}]\equiv\tilde\Gamma[\bar A]}$, with $\bar A$ any background configuration belonging to the background orbit $\bar {\cal A}$.\footnote{For this definition to make sense, $\tilde\Gamma[\bar A]$ should be the same for any background configuration $\bar A$ belonging to the same orbit $\bar {\cal A}$. That this is true is a direct consequence of Eq.~(\ref{eq:bg_inv}).} In terms of $\tilde\Gamma[\bar {\cal A}]$, the discussion of symmetry breaking can be done as usual:\footnote{The discussion is actually slightly more subtle since there could still exist multiple minimizing orbits in the Wigner-Weyl phase, see below for one specific example. What matters in the end is that there exists one symmetry-invariant minimizing orbit that allows one to identify the Wigner-Weyl phase.} in the Wigner-Weyl phase there will be only one orbit that minimizes $\tilde\Gamma[\bar {\cal A}]$ and this orbit is necessarily invariant in the sense of Eq.~(\ref{eq:0}); in the Nambu-Goldstone phase, there will exist various orbits that minimize $\tilde\Gamma[\bar {\cal A}]$, connected to each other by the symmetry. In general, it might not be so easy to think in terms of orbits. At finite temperature and for backgrounds of the form (\ref{eq:conf}), however, a convenient way is to use the Weyl chambers discussed in Sec.~\ref{sec:Weyl}.

A more serious difficulty lies in the fact that the identification of the minima of $\tilde\Gamma[\bar A]$ as actual states of the system (in the limit of zero sources) is not direct. As we recall in the next subsection, this identification relies on the fact that the minima of $\tilde\Gamma[\bar A]$ correspond to self-consistent backgrounds $\bar A_s$ such that 
\beq
\bar A_s=A_{\rm min}[\bar A_s]\,.\label{eq:self}
\eeq 
Then, they are also minima (with respect to variations of $A$) of the Legendre transform $\Gamma_{\bar A_s}[A]$ and, as such, qualify as actual states of the system. Similarly, the free-energy can be obtained solely in terms of $\tilde\Gamma[\bar A]$. Indeed
\beq
F=\Gamma_{\bar A_s}[A_{\rm min}[\bar A_s]]=\Gamma_{\bar A_s}[\bar A_s]=\tilde\Gamma[\bar A_s]\,.
\eeq
It follows that one can in principle study all the thermodynamical properties from $\tilde\Gamma[\bar A]$ and its absolute minima, the self-consistent backgrounds. 

As we will also see below, however, the identification of the self-consistent backgrounds with the minima of $\tilde\Gamma[\bar A]$ relies on the independence of the free-energy with respect to the choice of $\bar A$:
\beq
0=\frac{\delta}{\delta\bar A}\Gamma[A_{\rm min}[\bar A],\bar A]\label{eq:inv}
\eeq
which one can see as a particular case of Eq.~(\ref{eq:z_ind}). As we have already discussed, this identity might be violated by approximations or the degree of modelling. This can potentially introduce artefacts in the study of the deconfinement transition using self-consistent backgrounds which are a priori not present in the approach that relies on center-symmetric backgrounds.

\subsection{Self-consistent backgrounds as minima of $\tilde\Gamma[\bar A]$}

Let us now recall why self-consistent backgrounds identify with the minima of $\tilde\Gamma[\bar A]$. Consider first a self-consistent background $\bar A_s$. We can write
\beq
\tilde \Gamma[\bar A_s] & = & \Gamma[\bar A_s,\bar A_s]\nonumber\\
& = & \Gamma[A_{\rm min}[\bar A_s],\bar A_s]\nonumber\\
& = & \Gamma[A_{\rm min}[\bar A],\bar A]\nonumber\\
& \leq & \Gamma[\bar A,{\bar A}]=\tilde\Gamma[\bar A]\,,\label{eq:31}
\eeq
where we have successively made use of Eqs.~(\ref{eq:def}), (\ref{eq:self}), (\ref{eq:inv}), the definition of $A_{\rm min}[\bar A]$, and again Eq.~(\ref{eq:def}). It follows that $\bar A_s$ is an absolute minimum of $\tilde\Gamma[\bar A]$. Reciprocally, consider a background configuration $\bar A_{\rm min}$ that minimizes $\tilde\Gamma[\bar A]$ and assume that there exists at least one self-consistent background $\bar A_s$. We can write
\beq
\Gamma[\bar A_{\rm min},\bar A_{\rm min}] & = & \tilde\Gamma[\bar A_{\rm min}]\nonumber\\
& \leq & \tilde\Gamma[\bar A_s]\nonumber\\
& = & \Gamma[\bar A_s,\bar A_s]\nonumber\\
& = & \Gamma[A_{\rm min}[\bar A_s],\bar A_s]\nonumber\\
& = & \Gamma[A_{\rm min}[\bar A_{\rm min}],\bar A_{\rm min}]\,,\label{eq:32}
\eeq
where we have successively made use of Eq.~(\ref{eq:def}), the definition of $\bar A_{\rm min}$, followed again by Eq.~(\ref{eq:def}) and Eqs.~(\ref{eq:self}), (\ref{eq:inv}). It follows that $\smash{A_{\rm min}[\bar A_{\rm min}]=\bar A_{\rm min}}$ which means that $\bar A_{\rm min}$ is self-consistent.\footnote{The previous results can be easily understood from the following geometrical picture. The collection of pairs $(A_{\rm min}[\bar A],\bar A)$ defines a submanifold of the space $(A,\bar A)$. If Eq.~(\ref{eq:inv}) holds, this submanifold represents a ``flat valley'' for $\Gamma[A,\bar A]$ seen as a functional of both $A$ and $\bar A$. In other words, the points in this submanifold are minima of $\Gamma[A,\bar A]$ in $(A,\bar A)$-space. Now, self-consistent backgrounds are just the crossings of this flat valley with the submanifold $(\bar A,\bar A)$. They are then minima of $\Gamma[A,\bar A]$ in $(A,\bar A)$-space satisfying $\smash{A=\bar A}$ and then minima of $\tilde\Gamma[\bar A]$. Reciprocally, if there exists at least one self-consistent background, that is one crossing of the flat valley with the submanifold $(\bar A,\bar A)$, then any minimum of $\tilde\Gamma[\bar A]$ has the same depth than this self-consistent background and it should then belong to the flat valley while being such that $\smash{A=\bar A}$. It is thus a self-consistent background.} 

It is important to stress that this last argument relies on the existence of at least one self-consistent background. If the family of background gauges $z_{\bar A}[A]$ is such that $z_{\bar A=0}[A]$ does not break color symmetry explicitly, then we know that $\smash{\bar A=0}$ is self-consistent because we do not expect color invariance to break spontaneously either, ensuring that $\smash{A_{\rm min}[\bar A=0]=0=\bar A}$. We shall comment further on this point below.

We have thus arrived at a formulation of the thermodynamical observables that relies only on the minimization of the functional $\tilde\Gamma[\bar A]$. This shows that, for the purpose of determining the free-energy and the related thermodynamical observables (including the transition temperature), working with $\tilde\Gamma[\bar A]$ is equivalent to working with $\Gamma[A,\bar A]$ and thus with $\Gamma_c[A]$. In particular, as long as center symmetry is not broken, one of the minima of $\tilde\Gamma[\bar A]$ should be located at the center-invariant configuration $\smash{\bar A=\bar A_c}$, just as the minimum of $\Gamma_c[A]$ should be located at $\smash{A=\bar A_c}$. Right above the transition, the minimum of $\tilde\Gamma[\bar A]$ and the minimum of $\Gamma_c[A]$ should simultaneously depart from $\bar A_c$ (since the self-consistent backgrounds are the minima of $\tilde\Gamma[\bar A]$) thus providing two {\it a priori} equivalent identifications of the transition. 

Coming back to the presence of a trivial self-consistent background $\smash{\bar A=0}$ at all temperatures, this seems in conflict with the ability of $\tilde\Gamma[\bar A]$ of probing the transition. Indeed, since there should be a minimum of $\tilde\Gamma[\bar A]$ at the non-center-symmetric configuration $\smash{\bar A=0}$ at all temperatures, one may wonder how it is possible to conclude to the presence of a center-symmetric phase at low temperatures. In fact, what matters is that, in this range of temperatures, there is another minimum $\bar A_s$ located at the center symmetric point $\bar A_c$. That $\smash{\bar A=0}$ is always a minimum of $\tilde\Gamma[\bar A]$ is expected since, at least in the absence of approximations or modelling, the free-energy is background independent and should be equally computable in the absence of background. This leads precisely to the fact that $\smash{\tilde\Gamma[\bar A=0]}$,  which is nothing but the free-energy $\Gamma[A_{\rm min}[\bar A=0],\bar A=0]$ as computed in the no-background gauge, needs to have the same depth as any other minimum of $\tilde\Gamma[\bar A]$. As we will see below, in the presence of approximations (which typically break (\ref{eq:inv})) and even though $\smash{\bar A=0}$ remains a self-consistent background, it is not anymore a minimum of $\tilde\Gamma[\bar A]$, which, in a sense, lifts any sort of ambiguity regarding the choice of minimum of $\tilde\Gamma[\bar A]$.

\subsection{In practice}\label{sec:practice}
The equivalence between $\Gamma[A,\bar A]$ and $\tilde\Gamma[\bar A]$ relies on the central identity (\ref{eq:inv}). However, as we have mentioned above, this identity is difficult to maintain in the presence of the approximations and/or the degree of modelling required by most approaches. We have already seen that this compromises the use of $\Gamma[A,\bar A]$ with an arbitrary choice of $\bar A$, and favours the choice $\smash{\bar A=\bar A_c}$. From Eqs.~(\ref{eq:31}) and (\ref{eq:32}), it becomes clear that this compromises the use of $\tilde\Gamma[\bar A]$ as well. Indeed, although $\tilde\Gamma[\bar A]$ shares similar properties as $\Gamma_c[A]$ with respect to center-symmetry, in particular its minima $\bar A_{\rm min}$ are also order parameters, the very rationale for using these minima heavily relies on (\ref{eq:inv}).\footnote{We stress here that the problem lies not only in whether $\Gamma[A_{\rm min}[\bar A_c],\bar A_c]$ gives the same result as $\Gamma[A_{\rm min}[\bar A_s],\bar A_s]$ but also in whether the self-consistent backgrounds $\bar A_s$ can be identified with the minima $\bar A_{\rm min}$ of $\tilde\Gamma[\bar A]$ in the first place.} This implies that, even though $\tilde\Gamma[\bar A]$ and $\Gamma_c[A]$ both provide favourable frameworks to identify the deconfinement transition, the results for the transition temperature could differ and the ones based on $\Gamma_c[A]$ should be more reliable.

To date, there is no continuum approach where (\ref{eq:inv}) is exactly fulfilled and thus where the practical use of $\tilde\Gamma[\bar A]$ is entirely justified.\footnote{Besides, were it justified, it would become equivalent to using $\Gamma_c[A]$.} For instance, there is no rigorous derivation of the Gribov-Zwanziger action for background gauges at finite temperature. There exist models, such as the one proposed in Ref.~\cite{Kroff:2018ncl}, which restrict to self-consistent backgrounds to provide effective descriptions of the Polyakov loop potential, but they do not necessarily ensure the background independence of the partition function.\footnote{For recent new developments, see Ref.~\cite{Dudal:2023nbt}.} The same is true for the background field Curci-Ferrari (CF) model for which $\delta \Gamma[A_{\rm min}[\bar A],\bar A]/\delta \bar A$ vanishes when evaluated for a self-consistent background but does not vanish a priori for an arbitrary background \cite{Reinosa:2015gxn}.\footnote{We are currently investigating a way to cure this problem within the CF model.} This also applies a priori to non-perturbative continuum approaches that take as their starting action the Faddeev-Popov action. The need for a mass subtraction term \cite{Aguilar:2008xm,Quandt:2015aaa,Cyrol:2016tym} in order to deal with quadratic divergences can potentially jeopardize the background independence of the partition function.

On the other hand, the approach that we are proposing here does not suffer from this problem. Indeed, being based on the use of a regular Legendre transform, it does not rely on (\ref{eq:inv}). Its predictions are then more reliable and, as we have argued above, potentially testable on the lattice. We mention, however, that the comparison with lattice results may be spoiled by other subtleties of the gauge fixing in the continuum. In particular, with the notable exception of the Gribov-Zwanziger approach which restricts the functional integrals to first Gribov region, the continuum gauge fixings very often do not guarantee the positivity of the gauge-fixed measure which can potentially spoil the relation between the limit of zero sources and the minimization of the effective action and thus the comparison to the lattice.\footnote{It should be stressed, however, that the expected generation of a gluon mass has the capability to suppress contributions beyond the first Gribov region and thus to maintain to some extent the relation between the limit of zero sources and the minimization of the effective action.} The situation on the lattice is also subtle due to the presence of Gribov copies that spoils the analyticity of the effective action \cite{Maas:2013sca} and could compromise the testing of certain continuum properties that rely on derivatives of the effective action. One such property is precisely the appearance of zero modes in the inverse SU($2$) propagator which in the continuum is nothing but the Hessian of the effective action. However, if the analyticity violations are mild enough, one may expect a signal at the transition, either as a (sharp) peak of the susceptibility \cite{Reinosa:2016iml}, or even a divergence located in the vicinity of the transition. Gribov copies could have even more pernicious effects when implementing symmetric gauge fixings on the lattice but we leave these considerations for a future study.

\subsection{Vertex functions in the self-consistent approach}\label{sec:vertex}
We have not yet discussed the vertex functions in the presence of self-consistent backgrounds. There is a good reason for this that we now would like to clarify now. The point is that the self-consistency condition (\ref{eq:self}) possesses various solutions and, if one wants to define a {\it self-consistent background gauge}, one needs to specify which solution is chosen (at each temperature). 

We have seen for instance that $\smash{\bar A=0}$ is always a solution to (\ref{eq:self}) in the case where the no-background gauge does not explicitly break color invariance. There are in fact many other solutions that one can construct. Indeed, given a particular solution $\bar A_s$ (which could be $\bar A=0$), one can generate infinitely many other solutions by considering $\bar A_s^U$ with $U\in {\cal G}$.\footnote{We have seen above that $A^U_{\mbox{\tiny min}}[\bar A]$ is a minimum of $\Gamma[A,\bar A^U]$. It follows that $\smash{\bar A_s^U=A^U_{\mbox{\tiny min}}[\bar A_s]}$ is a minimum of $\Gamma[A,\bar A_s^U]$ and thus $\bar A_s^U$ is also a solution to (\ref{eq:self}), as announced.} All the solutions $\bar A_s^U$ that stem from a given $\bar A_s$ could/should be associated to the same gauge since the corresponding correlation functions are trivially related to each other (as one deduces by following a similar argument as in Sec.~\ref{eq:no_bg} but with $U\in {\cal G}$).\footnote{In particular, in the phase with broken symmetry where $\bar A_s$ and $\bar A_s^U$ (with $U$ a non-trivial center transformation) represent distinct center vacua, these different versions of the correlation functions correspond to the vertex functions in each of those vacua, but should be considered as correlation functions in the same gauge.} However, there are various inequivalent families of solutions $\bar A_s^U$. In the absence of approximations, we expect typically two such families, the one stemming from $\smash{\bar A=0}$ and the one stemming from $\smash{\bar A=\bar A_c}$ at low temperatures. Since $\smash{\bar A=0}$ corresponds already to the no-background gauge, it seems natural to define the self-consistent background gauge using the other family of solutions to (\ref{eq:self}) which we denote $\bar A_s$ from now on.

Now, because this $\bar A_s$ becomes equal to $\bar A_c$ in the confining phase, this definition of the self-consistent background gauge coincides with the center-symmetric gauge in this phase, and any conclusion about a given vertex function in one gauge applies to the other gauge as well. For instance, the appearance of a zero-mode in the SU(2) two-point function at $T_c$ in the center-symmetric gauge (associated to the continuous nature of the transition in this case) should also be imprinted in the two-point function evaluated in the self-consistent background gauge. In contrast, in the deconfined phase, $\bar A_s\neq \bar A_c$ and the two gauges are different, so the vertex functions are expected to be different as well. 

The previous discussion is an idealization, however, since, in the presence of approximations or modelling, the situation is more delicate. Due to the non-exact fulfilment of (\ref{eq:inv}), there can appear more families of solutions to (\ref{eq:self}) which introduce an ambiguity when defining the self-consistent background gauge. Within the Curci-Ferrari model, for instance, it is found that only one such family corresponds to a minimum of $\tilde\Gamma[\bar A]$ and it is natural to choose it as our definition of the self-consistent background gauge. This is the choice that was made for instance in \cite{Reinosa:2016iml}. Nonetheless, because of the loss of (\ref{eq:inv}) there is no reason for the minima to be in one-to-one correspondance with the self-consistent backgrounds and it is found that, in the range $[\bar T_c,T_c]$, where $\bar T_c$ and $T_c>\bar T_c$ denote the transition temperatures as obtained respectively from $\tilde\Gamma[\bar A]$ and $\Gamma_c[A]$, while the minimum of $\tilde\Gamma[\bar A]$ has already moved away from $\bar A_c$, the configuration $\smash{\bar A=\bar A_c}$ is still a self-consistent background. This implies that the vertex functions in the explicitly symmetric gauge and in the self-consistent background gauge, although they should in principle coincide up until the transition, differ in practice in the range $[\bar T_c,T_c]$. This explains why the zero-momentum mass defined in \cite{Reinosa:2016iml} never reached zero whereas the one defined within the explicitly symmetric gauge vanishes \cite{vanEgmond:2021jyx}. This adds to the fact that the choice of the explicitly symmetric gauge should be more sound than that of the self-consistent Landau gauge.

 \subsection{Hosotani mechanism}\label{sec:Hosotani}
The self-consistent background framework is also the one used in Ref.~\cite{Pastor-Gutierrez:2022rac} to discuss the Hosotani mechanism. In Sec.~\ref{sec:gauge_break}, we have already explained that the classification of phases found in Table I of that reference, and based on a certain degree of breaking of gauge invariance, is different from the one that we put forward in this paper and according to which any gauge-field configuration is compatible with gauge invariance, in the sense that the invariance under a symmetry at the level of gauge-field configurations needs always to be defined modulo gauge transformations.

Nevertheless, the classification in Table I of Ref.~\cite{Pastor-Gutierrez:2022rac} has a nice interpretation in terms of the Weyl chambers, each phase corresponding to either the vertices, edges or interior of the Weyl chambers. The question is now whether a phase transition could occur between these various phases. We stress that this would not affect our interpretation about gauge symmetry not being broken but this could leave some imprint on the free-energy in the form of an irregularity. This question has been investigated within the Curci-Ferrari model at one- and two-loop order in the self-consistent background framework and also at one-loop order within the center-symmetric background framework.

At one-loop order of the self-consistent background approach, a transition from phase $C$ to phase $A$ is indeed found above the center-breaking transition. However, it is also found that this transition is washed out by two-loop corrections, whereas the center-breaking transition remains \cite{Reinosa:2015gxn}. This seems to indicate that the transition from phase $C$ to phase $A$ is an artefact of the one-loop calculation. The situation is even cleaner in the center-symmetric background framework where no trace of such type of transition is seen at one-loop order.

\subsection{Higgs mechanism}\label{sec:Higgs}
 In the case of the Higgs mechanism, the effective action within a given gauge writes $\Gamma[A,\phi]$. If the gauge is chosen not to break global gauge transformations, we may restrict to $\smash{A=0}$ and define $\smash{\Gamma[\phi]\equiv\Gamma[A=0,\phi]}$ with $\smash{\Gamma[U_0\phi]=\Gamma[\phi]}$ for any constant $U_0$, which, in this work, we interprete as part of the group ${\cal G}_0$ of gauge transformations.
 
 The question is now whether a non-zero minimum of $\Gamma[\phi]$ should be interpreted as the breaking of the symmetry, as it would in the case where $\Gamma[\phi]$ corresponds to the effective action of a physical spin system. As we already discussed in the classical case, the answer to this question is negative because, in the case of the gauge system, symmetries should always be thought modulo gauge transformations and, in this sense, any configuration, even one with $\smash{\phi\neq 0}$, is compatible with the gauge symmetry. 
 
 The difference with the usual case of a scalar theory associated to a physical spin system is that we have here no way to build a physical source term that would select one preferred direction for the non-zero $\phi$. They can thus be selected through a modification of the gauge fixing  but not through the use of physical source terms unlike what happens with physical spin systems. These distinct directions should be seen as possible gauge choices that would further reduce the global gauge freedom. This happens for example in the 't Hooft gauges, see Ref.~\cite{Maas:2017wzi}. 
 
One could still argue that a non-zero $\phi$ points to the presence of multiple vacua connected by global gauge transformations and thus to a Nambu-Goldstone realization of the symmetry. However, just as we did above for $\tilde\Gamma[\bar A]$, the symmetry $\smash{\Gamma[U_0\phi]=\Gamma[\phi]}$ allows one to define the effective action directly on the global orbits $\smash{\Phi=\{U_0\phi,\, \mbox{with}\, U_0\, \mbox{global}\}}$ as $\smash{\Gamma[\Phi]\equiv\Gamma[\phi]}$ for $\smash{\phi\in\Phi}$. From this perspective, there is always only one vacuum, as given by the global orbit $\Phi$ that minimizes $\Gamma[\Phi]$ and the symmetry is realized in the Wigner-Weyl sense. Similar conclusions have been obtained in Refs.~\cite{Hertzberg:2018kyi,Hertzberg:2019ffc} through explicit calculations, even though the premises slightly differ from the ones considered in this work (specially the fact that we consider color rotations as redundancies).

Another question that comes to mind is what is the nature of the quantum state $|\phi\rangle$ that would be associated to a minimum at $\smash{\phi\neq 0}$. Should one conclude that the system explores states which violate color neutrality? Not necessarily, as we now speculate. Indeed, for the very same reason that a given state is described in terms of a global orbit $\smash{\Phi=\{U_0\phi,\, \mbox{with}\, U_0\, \mbox{global}\}}$ that stems from $\phi$, what really would describe the corresponding quantum state is the collection of all $U_0|\phi\rangle$ stemming from that particular $|\phi\rangle$. In a certain sense, this would be a non-Abelian generalization of the notion of rays in Quantum Mechanics.\footnote{Note that the physics would not depend on the particular representative, $|\phi\rangle$ or $U_0|\phi\rangle$, chosen within the ray. This is because observables are represented by operators that commute with gauge transformations and in particular with global gauge transformations, $OU_0=U_0O$. It follows that
\beq
\langle\phi|U^\dagger_0 OU_0|\phi\rangle=\langle\phi|U^\dagger_0 U_0O|\phi\rangle=\langle\phi|O|\phi\rangle\,.\nonumber
\eeq
In this sense, the non-Abelian color charge states would not be observable but only the associated color representation. In the Abelian case, the charge state is observable since it coincides with the representation and can be associated directly to the rays.} No matter what phase the system would be in, the states would be color neutral for the associated rays are color rotation invariant by construction.\footnote{We are here considering a gauge that respects color rotation invariance. In a gauge that would break color rotation invariance, we would have access only to one representative within each ray. The other representatives would describe the same state in color-rotated versions of the original gauge.} The non-trivial question would be how the system could transition from a phase where the representatives of the ray are themselves color neutral, to a phase where they are not color neutral. 
 
 We leave these interesting questions together with a thorough investigation of the notion of non-Abelian rays for a future work. Let us mention, however that this notion also allows one to lift one ambiguity related to the definition of the Polyakov loop. The latter is gauge-invariant. Yet, it seems to be associated to a colored object, a static test quark. One way to avoid the paradox would be to associate the Polyakov loop to the corresponding non-Abelian ray which is color neutral even though its representatives are not. In other words, the Polyakov loop is not associated to a colored object within a particular color state but, rather, to the corresponding color representation.

\section{Conclusions}
Within the context of non-abelian gauge theories, we have investigated the impact of gauge fixing, which deals with the (unphysical) gauge symmetries of these theories, on possible other (physical) symmetries. This impact can take two different forms. First, in the presence of approximations, the symmetry constraints on observables are not explicit in all gauges. Second, with or without approximations, the physical symmetries are not necessarily explicit at the level of the gauge-fixed correlation functions. 

Using the example of center symmetry at finite temperature, we have analyzed the origin of these inconvenient features within a very large class of gauges that contains the well known cases. We have also introduced the notion of symmetric gauge fixings for which these problems are absent. In particular, within such gauges, the correlation functions reflect explicitely the physical symmetry under scrutiny and can then serve as order parameters for the potential breaking of the latter. We have shown how to explicitly construct symmetric gauge fixings from background gauge fixings and we have compared our approach to the standard one based on the use of self-consistent backgrounds, pointing out the main differences and emphasizing why our new proposal should be more robust.

Although peculiar to the finite temperature case (or more generally to situations that involve compact dimensions), center symmetry possesses already the general features that allow one to extend the present discussion to other physical symmetries in the presence of unphysical degrees of freedom. In this respect, the important notion is that of invariance modulo gauge transformations. Not only does it underlie the whole construction in the present work, but it also allows one to (re-)interprete some of the transitions that can occur within gauge theories without invoking the conceptually annoying ``breaking of a gauge symmetry''. This point has been illustrated using various examples.

\acknowledgements{We would like to thank C\'edric Lorc\'e, Jack Holguin, Julien Serreau and Matthieu Tissier for fruitful discussions related to the present work.}


\appendix

\section{Physical symmetries and gauge fields}\label{sec:sym}
Let us now see how the discussion in the main text extends to any physical symmetry, beyond the particular case of center symmetry.

\subsection{Symmetries on field space}
Take a formal symmetry group ${\cal T}$. Its action on the space of gauge-field configurations is defined in terms of a representation. This means that, to each $T\in{\cal T}$, one associates a transformation $A\mapsto A^T$ that reflects the group structure as 
\beq
(A^{T_1})^{T_2}=A^{T_2T_1}\,.
\eeq 
Without loss of generality, we can assume that the so-obtained group of transformations on field space is in one-to-one correspondance with the group ${\cal T}$, so that it can be identified with it. If this is not the case from the start, one can always replace ${\cal T}$ by its quotient with the (normal) subgroup associated with the identity transformation on field space and the correspondance then becomes one-to-one.

One example is of course the group of gauge transformations ${\cal G}_0$. The subgroup of ${\cal G}_0$ that maps onto the identity transformation in field space is easily seen to correspond to constant color rotations of the form $\smash{U_0=e^{i2\pi k/N}\mathds{1}}$.\footnote{Those should not be mistaken with center transformations.} In what follows, we assume that ${\cal G}_0$ has been quotiented by this subgroup.

\subsection{The case of physical symmetries}
In the case of a physical symmetry, the representation of ${\cal T}$ on field space should obey a few additional properties which we now identify.

Recall first that two gauge-field configurations $A$ and $A^{U_0}$ related by a gauge transformation $U_0\in {\cal G}_0$ represent one and the same physical state. A physical transformation $T$  should then be such that $A^T$ and $(A^{U_0})^T$ also represent the same state. Therefore, there should exist a gauge transformation $U_0'\in {\cal G}_0$ such that 
\beq
(A^{U_0})^T=(A^T)^{U_0'}\,.\label{eq:cond}
\eeq 
Equivalently, this means that the action of the symmetry group on field space can equally be seen as an action on the space of ${\cal G}_0$-orbits. This is expected because the latter represent the physical states and, therefore, a physical symmetry should act directly on them. In other words, to each $T\in {\cal T}$, one can associate a transformation ${\cal A}\mapsto {\cal A}^T$ on the space of ${\cal G}_0$-orbits.

The condition (\ref{eq:cond}) is however not enough to characterize a physical symmetry group.\footnote{In fact, the condition also applies to the transformations in ${\cal G}_0$: if $T$ is a gauge transformation, we can write $(A^{U_0})^T=A^{TU_0}=A^{TU_0T^\dagger T}=(A^T)^{U_0'}$ with $U_0'=TU_0T^\dagger$.} This is because, among the field space transformations $A\mapsto A^T$, there could be some, other than the one corresponding to the identity of ${\cal T}$, that coincide with gauge transformations $A\mapsto A^{U_0}$. In this case, the subgroup of ${\cal T}$ associated with the identity transformation on orbit space is non-trivial and, in order to identify the actual physical symmetry group, one needs to quotient ${\cal T}$ by this subgroup.

For instance, in the discussion in the main text, we started from a group $\smash{{\cal T}={\cal G}}$ that, although it obeys (\ref{eq:cond}), is not physical for it contains a non-trivial subgroup ${\cal G}_0$ whose associated transformations on orbit space are all the identity transformation. The physical symmetry group is then ${\cal G}/{\cal G}_0$ in this case.

\subsection{Extension of physical symmetries}
In the previous section, we followed a top-down approach: starting from a group of symmetries that obeyed the condition (\ref{eq:cond}), we identified the associated physical group of symmetries by removing any gauge redundancy. Sometimes, it is convenient, instead, to consider a bottom-up approach: starting from a physical symmetry group, one enlarges it to include all possible gauge redundancies.

The construction is again based on the property (\ref{eq:cond}). Since we quotiented ${\cal G}_0$ by its center, $U_0'$ is uniquely determined in terms of $A$, $U_0$ and $T$. In what follows, we assume that it depends only on $U_0$ and $T$ and not on $A$. It is shown in this case that $U'_0(U_0,T)$ is compatible with both the group structure of ${\cal G}_0$ and the group structure of ${\cal T}$. For instance
\beq
(A^{U_0})^{T_1T_2} & = & ((A^{U_0})^{T_2})^{T_1}\nonumber\\
 & = & ((A^{T_2})^{U_0'(U_0,T_2)})^{T_1}\nonumber\\
 & = & ((A^{T_2})^{T_1})^{U_0'(U_0'(U_0,T_2),T_1)}\nonumber\\
 & = & (A^{T_1T_2})^{U_0'(U_0'(U_0,T_2),T_1)}\,,
\eeq
and so
\beq
U_0'(U_0,T_1T_2)=U_0'(U_0'(U_0,T_2),T_1)\,.
\eeq
Moreover,
\beq
(A^{U_{01}U_{02}})^{T} & = & ((A^{U_{02}})^{U_{01}})^T\nonumber\\
& = & (((A^{U_{02}})^T)^{U_0'(U_{01},T)}\nonumber\\
& = & ((A^T)^{U_0'(U_{02},T)})^{U_0'(U_{01},T)}\nonumber\\
& = & (A^T)^{U_0'(U_{01},T)U_0'(U_{02},T)}\,,
\eeq
and so
\beq
U_0'(U_{01}U_{02},T)=U_0'(U_{01},T)U_0'(U_{02},T)\,.
\eeq
In this case, the general theory of semi-direct products of groups, allows one to enlarge the group ${\cal T}$ into the semi-direct product $\smash{{\cal G}\equiv{\cal T}\ltimes {\cal G}_0}$ whose group law is defined as
\beq
(T_1,U_{01})(T_2,U_{02})=(T_1T_2,(U'_0)^{-1}(U_{01},T_2)U_{02})\nonumber\\
\eeq
and follows from
\beq
(((A^{U_{02}})^{T_2})^{U_{01}})^{T_1} & = & (((A^{U_{02}})^{(U'_0)^{-1}(U_{01},T_2)})^{T_2})^{T_1}\nonumber\\
& = & (A^{(U'_0)^{-1}(U_{01},T_2)U_{02}})^{T_1T_2}\,.
\eeq
One situation where this type of extension might be useful is that for which a given gauge fixing $z[A]$ is not be invariant under the original physical group ${\cal T}$. After extending ${\cal T}$ into $\smash{{\cal G}\equiv{\cal T}\ltimes {\cal G}_0}$, and even though $z[A]$ is not invariant under ${\cal G}$, it could turn out to be invariant under certain representatives of each of the classes of ${\cal G}/{\cal G}_0$, that is invariant under ${\cal T}$ modulo ${\cal G}_0$.

Take for instance a background gauge in which the background is the vector potential $\smash{\vec{A}(\vec{x})=(\vec{x}\times\vec{B})/2}$, already discussed in the main text. The gauge fixing is not invariant under the group of translations, but it is invariant under translations modulo gauge transformations.

\section{Conditional gauge fixings}\label{sec:cond}

Here, we discuss the specific case of conditional gauge fixings as given by $\smash{\rho[A]=\delta(F[A])}$, see Eq.~(\ref{eq:dofA}).\footnote{We shall later discuss the inclusion of a possible weight functional $w[A]$.} The question that we would like to address is under which conditions a given gauge-fixing functional $F[A]$ leads to a functional $z[A]$, see Eq.~(\ref{eq:zofA}), that is symmetric in the sense of Eq.~(\ref{eq:20}). We refer to that particular type of conditional gauge fixings as {\it symmetric conditional gauge fixings.}

\subsection{Necessary condition}
From Eqs.~(\ref{eq:zofA}) and (\ref{eq:dofA}), we see that a necessary condition for Eq.~(\ref{eq:20}) to hold true is that $\delta(F[A])$ and $\delta(F[A^U])$ have the same support in field-space, with $U$ any of the transformations appearing in Eq.~(\ref{eq:20}). In other words,
\beq
\forall {\cal U}\in {\cal G}/{\cal G}_0\,,\,\,\, \exists U\in {\cal U}\,,\,\,\, F[A]=0\Rightarrow F[A^U]=0\,.\label{eq:cc}
\eeq
We stress that this condition imposes strong constraints on the looked-after gauge-fixing functionals $F[A]$ for it means that the set of gauge-field configurations obeying $\smash{F[A]=0}$ should be invariant under specific (that is not all) transformations $U$ in each class ${\cal U}\in {\cal G}/{\cal G}_0$.

We also note that, similarly to the discussion in the previous subsection, a criterion such that 
\beq
\forall U\in {\cal G}\,,\,\,\, F[A]=0\Rightarrow F[A^U]=0\label{eq:cond1}
\eeq
makes not much sense because it implies that, given a gauge-field configuration $A$ obeying the gauge-fixing condition, all the gauge-field configurations in the same ${\cal G}_0$-orbit obey the condition as well, in contradiction with the fact that $F[A]$ should fix the gauge (up to possible Gribov copies). The same is true for the requirement
\beq
\forall U\in {\cal G}-{\cal G}_0\,,\,\,\, F[A]=0\Rightarrow F[A^U]=0\,,
\eeq
which is easily seen to imply (\ref{eq:cond1}) and thus leads again to a dead end.

\subsection{Sufficient condition}
In the absence of Gribov copies, the condition (\ref{eq:cc}) is not only necessary but also sufficient for Eq.~(\ref{eq:20}) to hold true. Indeed, for any transformation $U$ appearing in Eq.~(\ref{eq:cc}), we have 
\beq
\rho[A^U]=\alpha({\cal A})\rho[A]\label{eq:prop}
\eeq
where $\alpha({\cal A})$ depends only on the ${\cal G}_0$-orbit to which $A$ belongs and does not vanish on any orbit. It follows that
\beq
z[A^U] & = & \frac{\rho[A^U]}{\int_{{\cal G}_0} {\cal D}U_0\,\rho[(A^U)^{U_0}]}\nonumber\\
& = & \frac{\rho[A^U]}{\int_{{\cal G}_0} {\cal D}U_0\,\rho[A^{U_0U}]}\nonumber\\
& = & \frac{\rho[A^U]}{\int_{{\cal G}_0} {\cal D}U_0\,\rho[(A^{U^{-1}U_0U})^U]}\nonumber\\
& = & \frac{\alpha({\cal A})\rho[A]}{\int_{{\cal G}_0} {\cal D}U_0\,\alpha({\cal A})\rho[A^{U^{-1}U_0U}]}\nonumber\\
& = & \frac{\rho[A]}{\int_{{\cal G}_0} {\cal D}U_0\,\rho[A^{U_0}]}=z[A]\,,
\eeq
where we have used that $A^{U^{-1}U_0U}$ is in the same ${\cal G}_0$-orbit as $A$ and we have performed the change of variables $U_0\to UU_0U^{-1}$.

\subsection{Gribov copies}
In the presence of Gribov copies, one needs to add extra conditions in order to find a necessary and sufficient set of conditions for Eq.~(\ref{eq:20}) to hold true.

Before finding this set of conditions, let us show the following useful result, which is in fact valid for any $\rho[A]$ satisfying (\ref{eq:6}): $\smash{z[A^U]=z[A]}$ iff Eq.~(\ref{eq:prop}) holds true, with $\alpha$ possibly depending on $U$ as well in addition to ${\cal A}$. We have already shown one side of the equivalence, so let us show the other one. Assume that $\smash{z[A^U]=z[A]}$. Then 
\beq
\rho[A^U]=\alpha(A,U)\rho[A]\label{eq:rho2}
\eeq
with 
\beq
 \alpha(A,U)\equiv \frac{\int_{{\cal G}_0} {\cal D}U_0\,\rho[(A^U)^{U_0}]}{\int_{{\cal G}_0} {\cal D}U_0\,\rho[A^{U_0}]}\,,
\eeq
which we still need to show to depend only on the orbit of $A$ and that it does not vanish on any orbit. To this purpose, we write 
\beq
\alpha(A^{U_0'},U) & = & \frac{\int_{{\cal G}_0} {\cal D}U_0\,\rho[((A^{U_0'})^U)^{U_0}]}{\int_{{\cal G}_0} {\cal D}U_0\,\rho[(A^{U_0'})^{U_0}]}\nonumber\\
& = & \frac{\int_{{\cal G}_0} {\cal D}U_0\,\rho[A^{U_0UU_0'}]}{\int_{{\cal G}_0} {\cal D}U_0\,\rho[A^{U_0U_0'}]}\nonumber\\
& = & \frac{\int_{{\cal G}_0} {\cal D}U_0\,\rho[A^{U_0UU_0'U^{-1}U}]}{\int_{{\cal G}_0} {\cal D}U_0\,\rho[A^{U_0U_0'}]}\nonumber\\
& = & \frac{\int_{{\cal G}_0} {\cal D}U_0\,\rho[A^{U_0U}]}{\int_{{\cal G}_0} {\cal D}U_0\,\rho[A^{U_0}]}=\alpha(A,U)\,,
\eeq
as announced. That $\alpha(A,U)$ does not vanish follows from the left condition in (\ref{eq:6}).

We can now determine a set of necessary and sufficient conditions for (\ref{eq:prop}). Clearly, if the latter holds true, then $\rho[A]$ and $\rho[A^U]$ have the same support which means that Eq.~(\ref{eq:cc}) holds true. With this in mind, we can expand the $\rho[A]$ and $\rho[A^U]$ distributions along a given orbit ${\cal A}=\{A^{U_0}\,|\,U_0\in {\cal G}_0\}$. We find\footnote{Here, $\delta(U_0-V_0)$ is a somewhat formal notation for the delta distribution on the group such that $\int_{{\cal G}_0} {\cal D}U_0 \delta(U_0-V_0)f[U_0]=f[V_0]$. A more explicit form of Eqs.~(\ref{eq:1})-(\ref{eq:2}) can be obtained by locally considering a chart $U_0(\theta)$ on the Lie group manifold and expressing $\rho[A^{U_0(\theta)}]$ as a sum of $\delta(\theta-\theta^{(i)}(A))$'s.}
\beq
\rho[A^{U_0}]=\sum_i \left|{\rm det}\,\left.\frac{\delta F[A^{U_0}]}{\delta U_0}\right|_{U_0^{(i)}(A)}\right|^{-1}\delta(U_0-U_0^{(i)}(A))\label{eq:1}\nonumber\\
\eeq
and
\beq
& & \rho[(A^{U_0})^U]=\sum_i \left|{\rm det}\,\left.\frac{\delta F[(A^{U_0})^U]}{\delta U_0}\right|_{U_0^{(i)}(A)}\right|^{-1}\nonumber\\
& & \hspace{4.0cm}\times\,\delta(U_0-U_0^{(i)}(A))\,.\label{eq:2}
\eeq
Using Eq.~(\ref{eq:prop}) once more, we find that the ratio
\beq
\frac{\left|{\rm det}\,\left.\frac{\delta F[(A^{U_0})^U]}{\delta U_0}\right|_{U_0^{(i)}(A)}\right|^{-1}}{\left|{\rm det}\,\left.\frac{\delta F[A^{U_0}]}{\delta U_0}\right|_{U_0^{(i)}(A)}\right|^{-1}}\label{eq:criterion2}
\eeq
equals $\alpha({\cal A},U)$ and thus does not depend on $i$. Reciprocally, if we assume that $\rho[A]$ and $\rho[A^U]$ have the same support and that the ratio (\ref{eq:criterion2}) does not depend on $i$, it is easily seen that $\rho[A]$ obeys (\ref{eq:prop}). 

We have thus found that the general criterion for a conditional gauge fixing $\smash{F[A]=0}$ to be symmetric is that 1) in each class ${\cal U}\in {\cal G}/{\cal G}_0$, one can find a transformation $U$ that leaves the set of solutions to $\smash{F[A]=0}$ globally invariant, and  2) that the ratio (\ref{eq:criterion2}) does not depend on $i$ in the case where Gribov copies are present. In this latter case, let us also recall that the implementation of the conditional gauge fixing as a gauge fixing on average can be more generally done using
\beq
\rho[A]=w[A]\delta(F[A])\,.
\eeq
In this case, the ratio to be considered is
\beq
\frac{w[(A^{U_0^{(i)}(A)})^U]}{w[(A^{U_0^{(i)}(A)})]}\frac{\left|{\rm det}\,\left.\frac{\delta F[(A^{U_0})^U]}{\delta U_0}\right|_{U_0^{(i)}(A)}\right|^{-1}}{\left|{\rm det}\,\left.\frac{\delta F[A^{U_0}]}{\delta U_0}\right|_{U_0^{(i)}(A)}\right|^{-1}}\,.\label{eq:criterion3}
\eeq
In particular, suppose  that one finds a gauge-fixing functional such that
\beq
F[A^U]=UF[A]U^\dagger\,,
\eeq
for a certain $U$ in each class ${\cal U}\in {\cal G}/{\cal G}_0$. Then
\beq
\frac{\delta F^a[(A^{e^{i\theta}U_0})^U]}{\delta\theta^b}={\cal U}^{aa'}\frac{\delta F^{a'}[A^{e^{i\theta}U_0}]}{\delta\theta^b}\,,
\eeq
where ${\cal U}^{ab}$ is the adjoint representation of $U$. Since ${\rm det}\,{\cal U}=1$, we have
\beq
{\rm det}\,\frac{\delta F^a[(A^{e^{i\theta}U_0})^U]}{\delta\theta^b}={\rm det}\,\frac{\delta F^a[A^{e^{i\theta}U_0}]}{\delta\theta^b}
\eeq
and the condition (\ref{eq:criterion2}) is satisfied. That (\ref{eq:cc}) applies is also easily checked. It follows that the gauge-fixing functional $z[A]$ associated to $F[A]$ is center-symmetric. One explicit example of such type of gauge fixing conditions is the {\it center-symmetric background Landau gauge \cite{vanEgmond:2021jyx}}
\beq
F[A]\equiv D_\mu[\bar A_c](A_\mu-\bar A_{c,\mu})
\eeq
with $D_\mu[\bar A]\equiv\partial_\mu-i[\bar A_\mu,\,\,]$ and $\bar A_c$ and center-symmetric background.

\subsection{Orbit-dependent transformations}\label{sec:odt}
As already mentioned above, the condition (\ref{eq:cc}) imposes strong restrictions on the possible gauge fixing functionals $F[A]$ that lead to a symmetric gauge fixing in the sense of (\ref{eq:20}). In fact, it constrains the space of solutions of the condition $\smash{F[A]=0}$ to be stable under certain transformations $U$ in each of the classes ${\cal U}\in {\cal G}/{\cal G}_0$. This means actually two things: first, the solutions that are found along one orbit ${\cal A}$ and along the transformed orbit ${\cal A}^{{\cal U}}$ are pairwise-related by the same transformation $U\in{\cal U}$, and second, if we choose two other orbits $\tilde{\cal A}$ and $\tilde{\cal A}^{\cal U}$, the corresponding solutions to $\smash{F[A]=0}$ are related yet by the same transformation $U$.

This condition should not be mistaken with the property
\beq
\forall {\cal U}\in {\cal G}/{\cal G}_0\,,\,\,\, \forall A\,|\,F[A]=0\,,\,\,\, \exists U(A)\in {\cal U}\,,\,\,\, F[A^U]=0\,.\label{eq:int}\nonumber\\
\eeq
Although it might look similar to (\ref{eq:cc}), the condition (\ref{eq:int}) is a mere consequence of the natural assumption that the gauge fixing condition $\smash{F[A]=0}$ should intersect each ${\cal G}_0$-orbit at least once, and, therefore, applies to any sensible gauge fixing condition, be it symmetric or not. 

There is however one case where (\ref{eq:int}) implies a generalized version of (\ref{eq:20}) that might allow one to make the symmetry explicit. In the absence of Gribov copies, one can indeed define an orbit dependent transformation $U({\cal A})$ (not uniquely defined) which connects the configuration in a given orbit to that in the transformed orbit. Under this transformation, we can again write
\beq
\rho[A^{U({\cal A})}]=\alpha({\cal A})\rho[A]\
\eeq
from which we deduce that $\smash{z[A^{U({\cal A})}]=z[A]}$. Since $z[A]$ is given by Eq.~(\ref{eq:one_copy}), this rewrites
\beq
& & \delta(F[A^{U({\cal A})}]) \, \left|{\rm det}\left.\frac{\delta F[(A^{U({\cal A})})^{U_0}]}{\delta U_0}\right|_{U_0=1}\right|\nonumber\\
& & =\,\delta(F[A]) \, {\rm det}\left|\left.\frac{\delta F[A^{U_0}]}{\delta U_0}\right|_{U_0=1}\right|\,.\label{eq:sym0}
\eeq
It follows that the gauge-fixed functional integrand is invariant under this orbit dependent transformation. 

However, two questions remain to be answered to qualify this as a symmetry. First, the Jacobian of the transformation is non-trivial a priori since $U({\cal A})$ changes as one considers infinitesimal variations of the gauge field that are not taken along the corresponding orbit. This would need to be investigated further. Another potential problem is that only the product of the two factors in Eq.~(\ref{eq:one_copy}) is invariant, and not each factor separately. In practice, these factors are interpreted very differently, as an integral over the Faddeev-Popov ghost and antighost fields on the one hand, and as an integral over the Nakanishi-Lautrup field on the other hand. The symmetry (\ref{eq:sym0}) might be very non-trivial in terms of these fields and fragile to approximations (not to mention that the rewriting neglects the presence of the absolute value around the determinant). Finally, let us mention that an orbit dependent transformation such as $U({\cal A})$ cannot be used {\it a priori} to derive constraints on the generating functionals $W[J]$/$\Gamma[A]$ or on the corresponding correlation/vertex functions. This is because the transformation depends on the dynamical field (through the corresponding orbit) which serves as an integration variable under the functional integral.

Interestingly enough, in the lattice implementation of conditional gauge fixings (with or without Gribov copies), it is precisely the property (\ref{eq:int}) and, in a certain sense, also an orbit-dependent transformation, that allows one to ensure that the symmetry constraints at the level of the observables are explicit. Indeed, in the case of a conditional gauge fixing, the lattice selects one copy per orbit. This allows one to define an orbit dependent transformation that connects the copy on a given orbit to the copy in the transformed orbit under a certain ${\cal U}\in {\cal G}$ and, thus to relate the selected links with compensating values of the Polyakov loop. For the same reason as above, however, even on the lattice, the condition (\ref{eq:int}) is not enough to imply symmetry constraints on the vertex functions.

\section{Other symmetric gauge fixings}\label{sec:more}

In the main text, we have explained how to construct center-symmetric gauge fixings by combining background gauges and the notion of center-symmetric backgrounds. Here, we would like to investigate other possibilities and see whether or not they are viable.

\subsection{Center averaging}
Consider a gauge fixing on average $z[A]$ which, for simplicity, we assume to be invariant under color rotations of the form $e^{i\theta^j t^j}$. A priori, there is no reason for $z[A]$ to be invariant under center transformations. We will now explain how, from $z[A]$, one can build another gauge fixing on average $\bar z[A]$ that is center-invariant.

Consider a center transformation $\smash{U_1\in {\cal U}_1}$ of associated center element $e^{i2\pi/N}$ and define
\beq
\bar z[A] & \equiv & \frac{1}{N}\sum_{k=0}^{N-1} z[A^{U_1^k}]\nonumber\\
& = & \frac{1}{N}\sum_{k=0}^{N-1} z_{U_1^k}[A]\,.
\eeq
That $\bar z[A]$ defines a gauge fixing on average follows from the fact that each $z_{U_1^k}[A]$ averages to $1$ along any ${\cal G}_0$-orbit, see the discussion in the main text.

Let us now investigate how the functional $\bar z[A]$ transforms under $U_1$. We have
\beq
\bar z[A^{U_1}] & = & \frac{1}{N}\sum_{k=0}^{N-1} z[A^{U_1^{k+1}}]\nonumber\\
& = & \frac{1}{N}\sum_{k=1}^N z[A^{U_1^k}]\nonumber\\
& = & \bar z[A]+\frac{1}{N}\Big(z[A^{U_1^N}]-z[A]\Big)\,.
\eeq
We thus see that, if we were able to arrange for $\smash{z[A^{U_1^N}]=z[A]}$, then we would have $\smash{\bar z[A^{U_1}]=\bar z[A]}$ and so, even though $\smash{z[A^{U_1}]\neq z[A]}$.\footnote{If $\smash{z[A^{U_1}]=z[A]}$, there is no need to construct $\bar z[A]$ which in fact coincides with $z[A]$.}

To achieve this, let us choose $U_1$ such that it transforms a given Weyl chamber into itself. In particular, it leaves the confining configuration $A_c$ in that Weyl chamber invariant. It is easily argued that $U_1^N\in {\cal G}_0$ leaves each element of the Cartan subalgebra invariant\footnote{It is useful to associate to each element of the Weyl chamber, the corresponding value of the Polyakov loop. In particular, the vertices of the Weyl chamber are associated to the various center elements $e^{i2\pi k/N}$. Now, under the transformation $U_1$, the Polyakov loop is multiplied by $e^{i2\pi/N}$ which allows one to characterize the corresponding transformation on the Weyl chamber. The same applies for $U_1^N$ and because the Polyakov loop does not change in this case, we know that the corresponding transformation leaves the vertices invariant. But this transformation is an isometry acting on a space of dimension $N-1$. If it leaves $N-1$ linearly independent vectors invariant, it has to leave all points of the Weyl chamber invariant.} and thus it is a color rotation of the form $e^{i\theta^j t^j}$. Since we have assumed that $z[A]$ is invariant under such rotations, we have $\smash{z[A^{U_1^N}]=z[A]}$ and $\bar z[A]$ is center-invariant as announced. In what follows, it will be convenient to take $U_1$ of the form \cite{Reinosa:2019xqq,in_prep}
\beq
U_1(\tau)=e^{i\frac{\tau}{\beta}s^jt^j}W\,,\label{eq:choice}
\eeq
with some particular choices of $s^j$ and $W$ whose specific form is not needed here.

\subsection{Generating functional}
As we have discussed in Sec.~\ref{sec:sym_gf}, the generating functionals $W_{\bar z}[J]$ and $\Gamma_{\bar z}[A]$ obey the symmetry constraints that reflect the center-symmetry of the system. The question is now whether these functionals allow one to discuss the breaking of the symmetry. 

To investigate this question, we first derive the relation between $W_{\bar z}[J]$ and $W_z[J]$. We write
\beq
e^{W_{\bar z}[J]} & \equiv & \int {\cal D}A\,\bar z[A]\,e^{-S_{YM}[A]+J\cdot A}\nonumber\\
& = & \frac{1}{N}\sum_{k=0}^{N-1} \int {\cal D}A\,z[A^{U_1^k}]\,e^{-S_{YM}[A]+J\cdot A}\nonumber\\
& = & \frac{1}{N}\sum_{k=0}^{N-1} \int {\cal D}A\,z[A]\,e^{-S_{YM}[A]+J\cdot A^{({U_1^k})^\dagger}}\nonumber\\
& = & \frac{1}{N}\sum_{k=0}^{N-1} e^{\frac{i}{g}J\cdot ({U_1^k})^\dagger\partial U_1^k}\int {\cal D}A\,z[A]\,e^{-S_{YM}[A]+J^{{U_1^k}}\cdot A}\nonumber\\
& = & \frac{1}{N}\sum_{k=0}^{N-1} e^{\frac{i}{g}J\cdot ({U_1^k})^\dagger\partial U_1^k}e^{W_z[J^{{U_1^k}}]}
\eeq
that is
\beq
W_{\bar z}[J]=\ln\left(\frac{1}{N}\sum_{k=0}^{N-1} e^{\frac{i}{g}J\cdot ({U_1^k})^\dagger\partial U_1^k}e^{W_z[J^{{U_1^k}}]}\right).
\eeq
Suppose now that $z[A]$ is translation invariant. Using the assumed invariance of $z[A]$ under color rotations of the form $e^{i\theta^j t^j}$, and the particular choice (\ref{eq:choice}), it is then argued that $\bar z[A]$ is also translation invariant. Moreover, if we take the source $J$ constant and along the diagonal part of the algebra, the same holds for $J^{U_1^k}$. In this case both $W_{\bar z}[J]$ and $W_z[J^{{U_1^k}}]$ scale like the volume in the infinite volume limit. Introducing
\beq
w_{\bar z}[J] & \equiv & \lim_{V\to\infty} \frac{W_{\bar z}[J]}{\beta V}
\eeq
and
\beq
w_z[J] & \equiv & \lim_{V\to\infty} \frac{W_z[J]}{\beta V}\,,
\eeq
we thus find
\beq
w_{\bar z}[J] = {\rm Max}_k\left(\frac{i}{g}J\cdot ({U_1^k})^\dagger\partial U_1^k+w_z[J^{U_1^k}]\right)
\eeq
where $J\cdot ({U_1^k})^\dagger \partial U_1^k$ does not involve an integration over spacetime.

\subsection{Discussion}
The question is now whether one can study the confinement/deconfinement transition using the gauge fixing identified in the previous section. Let us consider the simple case where the original gauge fixing is color rotation invariant and take the SU(2) group. We can choose $U_1=i\sigma_2 e^{-i\pi\tau/\beta\sigma_3}$ and then
\beq
w_{\bar z}[J_3] = {\rm Max}\left(w_z[J_3],2\pi J_3+w_z[-J_3]\right)\,.
\eeq
Since we are interested in the limit $J_3\to 0$, we need to compare the functions $w_z[J_3]$ and $\tilde w_z[J_3]=2\pi J_3+w_z[-J_3]$ in the vicinity of $J_3=0$. We have $w_z[0]=\tilde w_z[0]$. Moreover, since $z$ is color invariant, we have $w_z'[0]=0$ and $\tilde w'_z[0]=2\pi$. Thus, in the vicinity of $J_3=0$, we find
\beq
w_{\bar z}[J_3>0] & = & \tilde w_z[J_3]\nonumber\\
w_{\bar z}[J_3<0] & = & w_z[J_3]\,.
\eeq
This means that $w_{\bar z}[J_3]$ presents a cusp at $\smash{J_3=0}$ for any temperature (even at low temperatures) and $w'_{\bar z}[J_3=0^+]=2\pi$ while $w'_{\bar z}[J_3=0^+]=0$. This example illustrates that having a symmetric gauge fixing is not enough for the gluon one-point function to serve as a probe for the deconfinement transition.

We stress that this conclusion applies even in the case where the original gauge-fixing $z[A]$ is completed in the infrared using the Curci-Ferrari model. In constrast, similar type of modelling applied on the center-symmetric Landau gauges discussed in the main text does allow to identify the transition from the gluon one-point function \cite{vanEgmond:2021jyx}. This plays thus in favor of the center-symmetric Landau gauges.

\subsection{Yet another possibility}
Another possibility is to introduce the following functional
\beq
W[J]=\frac{1}{N}\sum_{k=0}^{N-1}\left[\frac{i}{g}J\cdot ({U_1^k})^\dagger\partial U_1^k+W_z[J^{{U_1^k}}]\right].
\eeq
Because $W_z[J=0]$ is the free-energy $F$ of the system, we have
\beq
W[J=0]=\frac{1}{N}\sum_{k=0}^{N-1}W_z[J=0]=F
\eeq
which then gives also access to the free-energy. Moreover
\begin{widetext}
\beq
W[J^{U_1}] & = & \frac{1}{N}\sum_{k=0}^{N-1}\left[\frac{i}{g}J\cdot ({U_1^{k+1}})^\dagger(\partial U_1^k)U_1+W_z[J^{{U_1^{k+1}}}]\right]\nonumber\\
& = & \frac{1}{N}\sum_{k=0}^{N-1}\left[\frac{i}{g}J\cdot ({U_1^{k+1}})^\dagger\partial U_1^{k+1}+W_z[J^{{U_1^{k+1}}}]\right]-\frac{1}{N}\sum_{k=0}^{N-1}\frac{i}{g}J\cdot U_1^\dagger\partial U_1\nonumber\\
& = & W[J]-\frac{1}{N}\sum_{k=0}^{N-1}\frac{i}{g}J\cdot U_1^\dagger\partial U_1\,,
\eeq
which is a symmetry constraint similar to (\ref{eq:symW}). In the last steps, we have used that $U_1^N$ is a color rotation of the form $e^{i\theta^jt^j}$ and we have again assumed that $z[A]$ is invariant under such rotations.\\

\end{widetext}

Consider again the SU(2) case and assume that $z$ is color invariant. After dividing by the volume in the infinite volume limit, we have
\beq
w[J_3] = \frac{1}{2}\left(w_z[J_3]+2\pi J_3+w_z[-J_3]\right)\,.
\eeq
Now, because $w_z[J_3]$ has no cusp at $J_3=0$ since $w'_z[0]=0$, neither has $w[J_3]$, and we find
\beq
w'[0]=\frac{1}{2}(w'_z[0]+2\pi-w'_z[0])=\pi\,.
\eeq
This time the minimum is always in the symmetric configuration and thus cannot serve as a probe for the deconfinement transition either.

\end{document}